\documentstyle[12pt,epsf,amsfonts,amssymb]{article}

\def\ben{\begin{equation}}
\def\een{\end{equation}}
\def\bena{\begin{eqnarray}}
\def\eena{\end{eqnarray}}
\def\non{\nonumber}
\def\D{D}
\def\E{{\cal E}}
\def\scalar{{\Bbb S}}
\def\vector{{\Bbb V}}
\def\tensor{{\Bbb T}}
\renewcommand{\H}{{\mathcal H}}
\newcommand{\tg}{{\tilde g}}
\newcommand{\tnabla}{{\tilde \nabla}}
\newcommand{\tn}{{\tilde n}}
\newcommand{\K}{{K}}
\newcommand{\R}{{\mathcal R}}
\newcommand{\Q}{{\mathcal Q}}
\newcommand{\I}{{\mathscr I}}

\DeclareFontFamily{U}{rsfs}{\skewchar\font"7F}
\DeclareFontShape{U}{rsfs}{m}{n}{<-6> rsfs5 <6-8> rsfs7<8-> rsfs10 }{}
\DeclareMathAlphabet{\mathscr}{U}{rsfs}{m}{n}

\author{Stefan Hollands\thanks{\tt hollands@vulcan2.physics.ucsb.edu} \\
{\it Physics Department, UCSB, Santa Barbara, CA 93106, USA, and}\\
{\it Inst. f. Theor. Phys., G\" ottingen U, D-37077 G\" ottingen, Germany,}\\
\\
Akihiro Ishibashi\thanks{\tt akihiro@midway.uchicago.edu}\\
{\it Enrico Fermi Inst., U Chicago, Chicago, IL 60637, USA}\\
\\
Donald Marolf\thanks{\tt marolf@physics.ucsb.edu}\\
{\it Physics Department, UCSB, Santa Barbara, CA 93106, USA} }

\title{Comparison between various notions of conserved charges in
  asymptotically AdS-spacetimes}

\begin{document}

\maketitle
\begin{abstract}
We derive hamiltionian generators of asymptotic symmetries for general relativity
with asymptotic AdS boundary conditions using the ``covariant phase space'' method
of Wald et al. We then compare our results with other definitions that have been
proposed in the literature. We find that our definition agrees with that proposed by Ashtekar
et al, with the spinor definition, and with the background dependent definition of Henneaux
and Teitelboim. Our definition disagrees with the one obtained from the ``counterterm subtraction
method,'' but the difference is found to consist only of a ``constant offset'' that is
determined entirely
in terms of the boundary metric. We finally discuss and justify our boundary conditions by a linear
perturbation analysis, and we comment on generalizations of our
boundary conditions, as well as inclusion of matter fields.
\end{abstract}

\section{Introduction}

While the cosmological constant observed in nature seems to be
(small and) positive, there is considerable theoretical interest
in studying theories of gravity with a negative cosmological
constant. A negative cosmological constant gives rise to classical
spacetimes whose asymptotic structure can be viewed as a timelike
``boundary'' at infinity, which in turn can be considered as a
lower dimensional spacetime in its own right. The AdS/CFT
correspondence~\cite{Juan,GKP,Witten,MAGOO} asserts that conformal
quantum field theories living on this boundary provide a
``holographic'' description of the gravity theory in the ``bulk''.

An important aspect of the correspondence is the matching of symmetries on both sides.  In diffeomorphism invariant theories such as general relativity in $d$ dimensions, the hamiltonian generators of asymptotic spacetime symmetries can be expressed as
surface integrals over a $(d-2)$-dimensional cross section at infinity, while the asymptotic symmetry
vector fields themselves can be identified with conformal killing fields of the boundary. On
the field theory side, these generators correspond to generators of the conformal
symmetry algebra of the CFT, which are also given by integrals over
$(d-2)$ dimensional surfaces within the boundary
spacetime\footnote{The same statement is also true for the generators
of the ``fermionic'' symmetries that occur in supergravity. We will restrict ourselves here to
the bosonic generators.}.

Given the importance of symmetries for the correspondence, it is perhaps surprising that
our understanding of the hamiltonian generators of asymptotic symmetries in
asymptotically AdS spacetimes does not seem to be
optimal---as may be appreciated e.g. from that fact that, over the years, several
definitions for ``conserved charges'' associated with asymptotic symmetries
have been proposed: The definition by Ashtekar et al.~\cite{asht1, asht2} based on the electric Weyl tensor, the Hamiltonian definition by Henneaux and Teitelboim~\cite{ht}, the ``pseudotensor'' approach of Abbott and Deser~\cite{ad}, the KBL approach \cite{KBL,KBLapp}, the spinor definition~\cite{gary, warner} based on an original idea by Witten~\cite{w}, and the ``counterterm subtraction method''  by Henningson and Skenderis and by Balasubramanian and Kraus~\cite{skenderis,kraus,KS2,KS3,KS4,KS5,KS6,KS7,KS8,KS9}. These constructions are all rather different in philosophy and appearance. Their relation has mostly been analyzed in the context of special solutions to the field equations, but a comprehensive and systematic comparison does not seem to exist in the literature. The aim of the
present paper is to fill in this gap.

However, before doing so, we will begin in sections~2 and~3
by providing yet another construction of conserved charges
in asymptotically AdS-spacetimes.  For this purpose we use the general ``covariant phase space formalism'' of Wald et al.~\cite{wz, wi} (see also \cite{ABR} for earlier related work). Our construction combines, in some sense, many individual advantages of the previous
constructions, while avoiding some of their disadvantages. For example, it has the
advantage of being couched in a Hamiltonian framework (as does
the approach of Henneaux and Teitelboim), which ensures that the
charges have the proper physical interpretation as the generators
of the symmetries. At the same time, the construction
is manifestly ``covariant'' and ``background
independent'' because it uses the formalism of conformal infinity introduced
by Penrose~\cite{p} (as in Ashtekar's definition). Our construction will also facilitate the comparison with
various previous definitions of conserved charges.

This comparison is carried out in detail in section~4.  It turns out that the final form of our expression for
the conserved charges is manifestly equivalent to  Ashtekar's (although our derivation is different). In subsection~4.1, we compare our definition
with the counterterm subtraction method in $d=5$. We show that the methods are not equivalent, as
had in fact been noticed previously~\cite{kraus, asht2}. However, we show that the difference is given only by a
constant offset, which is expressible
entirely in terms of the non-dynamical boundary data, and hence is
the same for {\em any} asymptotically AdS spacetime. In subsection~4.2,
we compare our method to that of Henneaux and Teitelboim, and we find that both are equivalent.
In subsection~4.3, we establish that our definition is equivalent to the spinor method in $d$ dimensions, thereby generalizing a result of Davis~\cite{davis} in $d=4$.  However, comparison with the Abbott-Deser approach \cite{ad} and its generalizations  \cite{SD2,SD3,OK}, as well as with the KBL approach \cite{KBL,KBLapp}, is postponed for future investigation.
In section~5, we discuss how to generalize our constructions when matter fields are present. Of particular interest are scalar fields saturating the so-called ``Breitenlohner-Friedman bound''~\cite{bf}, and abelian $p$-form fields. These are therefore discussed in some detail in sections~5.1
and~5.2--5.4,  respectively.

Finally, in section~6 we motivate our choice of ``boundary conditions'' by showing that they arise naturally
in a linear perturbation analysis around exact AdS-space. In fact, that analysis shows that other
boundary conditions are also possible, and we discuss some of these,
although we leave a more detailed analysis, as well as the derivation
of the corresponding conserved charges to a future investigation.

\medskip

Our {\bf notations and conventions} are the following: The dimension of the spacetime
is denoted $d$, and we assume\footnote{
Since the Weyl tensor vanishes for $d=3$ the definition \cite{asht1,asht2} of Ashtekar et al becomes trivial for this case.  However, for $d=3$ a covariant phase space construction of energy has already been given in  \cite{AWD} and one has the Henneaux-Teitelboim type construction~\cite{BH} by Brown and Henneaux.
While it would be interesting to compare these latter definitions with the counter-term subtraction approach, the $d=3$ case requires special treatment due to the appearance of factors of $d-3$ in many formulae below.    We therefore restrict to $d \ge 4$ below, referring the reader to \cite{KS3} for details of the expansion in $d=3$ and saving the comparison of charges for \cite{peierls}.} $d \ge 4$. The signature
of the metric is $(-+++ \dots)$, the convention for the Riemann tensor
is $\nabla_{[a} \nabla_{b]} k_c = (1/2) {R_{abc}}^d k_d$
and $R_{ab} = {R_{acb}}^c$ for the Ricci tensor.
Indices in parenthesis are symmetrized and indices
in brackets are antisymmetrized. Indices on tilde tensor fields $\tilde t_{abc\dots}$
are raised and lowered with the unphysical metric $\tilde g_{ab}$ and its inverse $\tilde g^{ab}$,
whereas indices on untilde fields are raised and lowered with the physical metric $g_{ab} = \Omega^{-2}
\tilde g_{ab}$, and its
inverse $g^{ab}$. The AdS radius is denoted $\ell$. We set $\ell = 1$ in most of our formulas
if not explicitly stated otherwise.

\section{Hamiltonian approach to conserved charges in AdS
and boundary conditions}
\label{approach}

In this section, we review the general algorithm given by Wald and
Zoupas~\cite{wz} for defining ``charges'' associated with symmetries
preserving a given set of ``boundary conditions'' in the context of theories
derived from a diffeomorphism covariant Lagrangian. This will be used to
define conserved charges in $d$-dimensional general relativity with a specific
choice of asymptotic
AdS boundary conditions, as the generator conjugate to an appropriately
defined asymptotic symmetry.

The algorithm~\cite{wz} applies to arbitrary theories derived from a
diffeomorphism covariant Lagrangian. We focus here on vacuum
general relativity with a negative cosmological constant in
$d$ dimensions, defined by the Lagrangian density (viewed as a $d$-form)
\begin{equation}
\label{einsteinhilbert}
L = \frac{1}{16\pi G} \, \sqrt{ - g} \, (R - 2 \Lambda) \, d^d x,
\quad \Lambda < 0.
\label{lagrangian:Einstein}
\end{equation}
In order to completely specify the theory, one must also prescribe
a set of asymptotic conditions for the metric. In this paper,
we will consider the following

\paragraph{\bf Asymptotic Conditions:}
\begin{enumerate}
\item One can attach a boundary, $\I \cong R \times S^{d-2}$
to $M$ such that $\tilde M = M \cup \I$ is a manifold
with boundary.
\item
On $\tilde M$, there is a smooth\footnote{By ``smooth'', we mean
$C^\infty$. However, for our constructions to work, it would be sufficient
to require only that $\tilde g_{ab}$ is $(d-1)$-times continuously differentiable. This weaker
requirement is, in fact, the appropriate one when various matter fields are included. It would then
be natural to also weaken the differentiability of $\Omega$ and the manifold structure of $\tilde M$,
but we will not discuss this for simplicity.}
metric $\tg_{ab}$ and a smooth
function $\Omega$ such that $g_{ab} = \Omega^{-2} \tg_{ab}$, and such that $\Omega = 0$ and
\ben
\tn_a \equiv \tnabla_a \Omega \neq 0
\een
at points of $\I$. The metric $\tilde h_{ab}$ on $\I$ induced by $\tg_{ab}$
is in the conformal class of the Einstein static universe,
\ben
\tilde h_{ab} \, dx^a dx^b = e^{\omega} [-dt^2 + d \sigma^2],
\een
where $d\sigma^2$ is the line element of the unit sphere $S^{d-2}$,
and where $\omega$ is some smooth function. Thus, $\I$ is
a timelike boundary.
\end{enumerate}
Other boundary conditions, corresponding to different
notions of asymptotically AdS spacetimes are also possible (see
Sec.6), but will not be discussed here.
The prototype spacetime satisfying the above asymptotic conditions is,
of course, AdS space itself. It has the line element
\ben
\label{pureads}
ds^2_0 = - (1 + r^2/ \ell^2)\, dt^2 + \frac{dr^2}{1 + r^2/\ell^2}
+ r^2 d\sigma^2,
\een
in global coordinates, where
\ben
\ell
\equiv  \sqrt{-(d-1)(d-2)/2\Lambda}
\een
is the AdS radius.
By a change of coordinates, this can be brought into the form
\ben
ds_0^2 = \frac{\ell^2}{\Omega^2} \Bigg[ d\Omega^2 - dt^2 + d\sigma^2
                                -\frac{\Omega^2}{2}(dt^2 + d\sigma^2)
                                  + \frac{\Omega^4}{16}(-dt^2 + d\sigma^2)
                                 \Bigg]\,,
\label{ds00}
\een
where $\Omega > 0$ is an analytic function of $r$.
Thus, for pure AdS, a conformal completion can be obtained
by taking $\tilde M$ to be the manifold obtained from $M$ by attaching
the boundary $\I$ consisting of the points $\Omega=0$, and
by taking the unphysical metric to be $d\tilde s_0^2
= \Omega^2 ds^2_0$. This is explicitly
\ben
d \tilde s^2_0 = \ell^2 \Bigg[ d\Omega^2 - dt^2 + d\sigma^2
                               -\frac{\Omega^2}{2}(dt^2 + d\sigma^2)
                               + \frac{\Omega^4}{16}(-dt^2 + d\sigma^2)
                        \Bigg] \,,
\label{unphysads}
\een
and obviously smooth. The induced metric on $\I$
is $\ell^2 [-dt^2 + d\sigma^2]$, i.e.,
the metric of the Einstein static universe with radius $\ell$.
One can likewise verify that the asymptotic conditions are also
obeyed by the AdS-Schwarzschild and AdS-Myers-Perry solutions.

If the above asymptotic conditions are combined with Einstein's equation,
then one can obtain much more detailed results about the asymptotic form of the metric
at infinity. These consequences will be worked out in the next section and are summarized in
lemma~3.1.

The diffeomorphisms $f$ of $\tilde M$ with
the property that $f^* g_{ab}$ is asymptotically AdS whenever $g_{ab}$ is, form a
group under composition. Of physical significance is the
group $G$ obtained by factoring this group by $\mbox{Diff}(M)_0$,
where $\mbox{Diff}(M)_0$ is the subgroup of diffeos leaving $\I$ pointwise  invariant.
The factor group $G$ is called the ``asymptotic symmetry group''.
Since the elements of $G$ can be identified with conformal isometries of the Einstein
static universe, it follows that
\ben
G \cong O(d-1, 2).
\een
The elements of the Lie-algebra of $G$ correspond to equivalence classes
of vector fields $\xi^a$ generating a 1-parameter group asymptotic
symmetries, modulo vector fields that vanish on $\I$.
(By abuse of language, we will refer to these again
as ``asymptotic symmetries.'')
We are interested in defining the corresponding generators $\H_\xi$ on phase space\footnote{
Note that, since the Lie-algebra elements of $G$ are in correspondence
with equivalence classes vector fields on $\tilde M$ modulo vector fields
that vanish on $\I$, it follows that the expression for $\H_\xi$ must be such that
it is independent on which particular vector field representative in
the equivalence class is chosen.
Thus, it must vanish for any vector field that is zero on $\I$. This means, roughly speaking,
that $\H_\xi$ cannot depend on derivatives of $\xi$ normal to $\I$.}.

For this, consider the  variation of the Lagrange density $L$,
which can always be written in the form
\begin{equation}
\delta L = F \cdot \delta g + d \theta,
\end{equation}
Here, $F$ are the field equations; in our case
\begin{equation}
F_{ab} = \frac{1}{16\pi G} d^dx \sqrt{-g} \,
                    \left(
                       R_{ab} - \frac{1}{2} R g_{ab} + \Lambda g_{ab}
                    \right) \, ;
\end{equation}
and $d\theta$ is the exterior differential of the $(d-1)$-form $\theta$
corresponding to the ``boundary term'' that would arise if the variation of
$L$ were performed under an integral sign. It is given in our case by
\begin{equation}
\label{td}
\theta_{a_1 \dots a_{d-1}} = \frac{1}{16\pi G} v^c \epsilon_{ca_1
\dots a_{d-1}},
\end{equation}
where $\epsilon = d^d x \sqrt{-g}$ is the volume form (identified
with a $d$-form), and $v_a$ is given by
\ben
\label{vadef}
v^a = \nabla^b \delta g_b{}^a - \nabla^a \delta g_b{}^b.
\een
The antisymmetrized second
variation\footnote{Here, and in similar other formulas involving
second variations, we assume without loss of
generality that the variations
commute, i.e., that $\delta_1(\delta_2 g) - \delta_2(\delta_1 g) = 0$.}
$\omega$ of $\theta$ defines the (dualized) symplectic current,
\begin{equation}
\label{dw}
\omega(g; \delta_1 g, \delta_2 g)
= \delta_1 \theta(g; \delta_2 g) - \delta_2 \theta(g; \delta_1 g),
\end{equation}
so that $\omega$ depends on the unperturbed metric and is skew in the pair of
perturbations $(\delta_1 g, \delta_2 g)$.
It is given in our case by
\begin{equation}
\omega_{a_1 \dots a_{d-1}} = \frac{1}{16\pi G} w^c \epsilon_{ca_1 \dots a_{d-1}},
\end{equation}
where $w^c$ is the symplectic current vector
\begin{equation}
\label{wdef}
w^a = P^{abcdef}(\delta_1 g_{bc} \nabla_d \delta_2 g_{ef}
- \delta_2 g_{bc} \nabla_d \delta_1 g_{ef})
\end{equation}
with
\begin{equation}
P^{abcdef}
  = g^{ae} g^{fb} g^{cd}
  - \frac{1}{2} g^{ad} g^{be} g^{fc}
  - \frac{1}{2}g^{ab} g^{cd} g^{ef}
  - \frac{1}{2}g^{bc} g^{ae} g^{fd}
  + \frac{1}{2}g^{bc} g^{ad} g^{ef}.
\end{equation}
The integral of the symplectic current over an achronal
$(d-1)$-dimensional submanifold $\Sigma$ of $\tilde M$ defines the
symplectic structure, $\sigma_\Sigma$, of general relativity
\begin{equation}\label{symplectic}
\sigma_\Sigma(g; \delta_1 g, \delta_2 g) = \int_{\Sigma} \omega(g; \delta_1 g, \delta_2 g) .
\end{equation}
It follows from a general argument that when both $\delta_1 g$ and
$\delta_2 g$ satisfy the linearized equations of motion, then
$d\omega = 0$, or, what is the same thing, that the symplectic
current~(\ref{wdef}) is conserved, $\nabla^a w_a = 0$. This fact
can be used to find how $\sigma_\Sigma$ depends upon the choice of
$\Sigma$. Namely, let $\Sigma_1$ and $\Sigma_2$ be two achronal
surfaces ending on $\I$. They enclose a spacetime volume which is
bounded by $\Sigma_1, \Sigma_2$, and a portion $\I_{12}$ of
infinity. By Stokes theorem, the difference between $\sigma_1$ and
$\sigma_2$ is given by the integral $\int_{\I_{12}} \omega$.
However, below equation (\ref{lieeta}) we will show that, when
viewed as a form on the unphysical spacetime, the pull-back of
$\omega$ to $\I$ vanishes for linearized solutions $\delta_1 g$
and $\delta_2 g$ satisfying our asymptotic conditions.
Consequently, the symplectic structure $\sigma_\Sigma$ does not
depend on the choice of $\Sigma$.  Furthermore, since $\omega$
vanishes at $\I$ and $\Sigma$ is compact in the the unphysical
spacetime, this will show that the symplectic structure
(\ref{symplectic}) is well-defined. An equivalent---and somewhat
more familar---way to write the symplectic form is \ben
\sigma_\Sigma(g; \delta_1 g, \delta_2 g) = \frac{1}{16\pi G}
\int_\Sigma [\delta_1 \pi^{ij} \delta_2 q_{ij} - (1
\leftrightarrow 2) ] \een where $q_{ij}$ is the intrinsic metric
on $\Sigma$, and $\pi^{ij} = d^{d-1} x \sqrt{-q}(\kappa^{ij} -
\kappa q^{ij})$ the momentum, written in terms of the extrinsic
curvature $\kappa_{ij}$ of $\Sigma$.

One would now like to define the generator associated with a vector field $\xi^a$
representing an asymptotic symmetry by
\ben
\label{hvf0}
\delta \H_\xi = \sigma_\Sigma(g; \delta g, \pounds_\xi g) \quad \forall \delta g,
\een
where $\Sigma$ is a partial Cauchy surface whose boundary, $C$, is a cut of $\I$.
Note that, since the right side is the symplectic form, this equation says
that $\H_\xi$---if it exists---is
indeed the generator (in the sense of Hamiltonian mechanics)
of the infinitesimal displacement
(``Hamiltonian vector field'') $\delta g = \pounds_\xi g$,
which in turn describes the action of an infinitesimal symmetry in the phase space
of the theory.

Note that since
$\pounds_\xi \pounds_\eta - \pounds_\eta \pounds_\xi = \pounds_{[\xi,
  \eta]}$ (with $[\xi, \eta]^a
= \xi^b \nabla_b \eta^a - \eta^b \nabla_b \xi^a$
the commutator of two vector fields), it
follows automatically that the Hamiltonian vector fields associated with two infinitesimal symmetries $\xi^a, \eta^a$
satisfy the same algebra as ordinary vector fields on $M$ under the commutator. Consequently,
if the symplectic form $\sigma_\Sigma$
is used to define a Poisson bracket, then the charges $\H_\xi$---if they exist---must satify the same algebra up to a central extension, i.e.,
\ben
\label{pbracket}
\{\H_\xi, \H_\eta\} = \H_{[\xi, \eta]} + c(\xi,\eta).
\een
Note also that at this stage $\H_\xi$, $\H_\eta$, and $c(\xi, \eta)$ might perhaps depend on the choice of partial Cauchy surface $\Sigma$.

Before analyzing the existence of $\H_\xi$ within
the specific setup under investigation in this paper---namely
Einstein gravity with a negative cosmological constant and the boundary
conditions spelled out above---it is instructive, following~\cite{wz},
to first study the general structure of eq.~(\ref{hvf0}).
Let us assume that the the right side of eq.~(\ref{hvf0})
is actually finite, as will be the case, e.g., if the symplectic current form
$\omega(g, \delta_1 g, \delta_2 g)$ has a well-defined
(i.e., finite) extension to $\I$, and as will turn out to be the case in our
setup. Equation~(\ref{hvf0}) can then be written in the form~\cite{wz}
\ben
\label{hvf}
\delta \H_\xi = \int_\Sigma \delta {\mathcal C}_a \xi^a + \int_C [\delta Q_\xi - \xi \cdot \theta(g, \delta g)],
\een
Here, ${\mathcal C}_a = {\mathcal C}_{a[b_1 \dots b_{d-1}]}$
are the constraints of the theory\footnote{In the case of pure gravity
that we are considering here, these are given by ${\mathcal C}_{a[b_1
\dots b_{d-1}]}= \epsilon_{b_1 \dots b_{d-1}c}(R_a{}^c - \frac{1}{2} R
\delta_a{}^c + \Lambda \delta_a{}^c)$.} (identified with $(d-1)$-forms),
and $Q_\xi$ is the Noether charge, which is given in our case by
\ben
\label{nc}
Q_{a_1 \dots a_{d-2}} = -\frac{1}{16\pi G}(\nabla^b \xi^c) \epsilon_{bca_1 \dots a_{d-2}}\,.
\een
Consistency
requires $(\delta_1 \delta_2 - \delta_2 \delta_1) \H_\xi = 0$, so we must have
\ben
\label{consistency}
0 =
\xi \cdot [\delta_2 \theta(g, \delta_1 g) - \delta_1 \theta(g, \delta_2 g)] =
\xi \cdot \omega(g; \delta_1 g, \delta_2 g)
\een
on $\I$, or else $\H_\xi$ cannot exist. If this equation holds---which is by no means guaranteed
in general and depends crucially on the Lagrangian {\em and} on the nature of the boundary conditions---then
it follows\footnote{In order to prove this statement from the consistency condition~(\ref{consistency}),
one needs to assume that the space of asymptotic AdS-geometries is
simply connected~\cite{wz}.} that there is a $(d-2)$-form $I_\xi$ such that
\ben
\delta Q_\xi - \xi \cdot \theta(\delta g) = \delta I_\xi
\een
up to an exact form. We conclude that a solution to eq.~(\ref{hvf}) is given by
\ben
\label{hdef1}
\H_\xi = \int_\Sigma \xi^a {\mathcal C}_a + \int_C I_\xi.
\label{db}
\een
When the equations of motion are satisfied, then the constraints ${\mathcal C}_a$ vanish identically, so
the first term vanishes and $\H_\xi$ reduces to a surface integral.

As we will see, the consistency condition~(\ref{consistency}) holds
(in fact, the symplectic form $\omega$ vanishes identically
on $\I$) under our asymptotic conditions (in combination with Einstein's equations),
and so a conserved generator
$\H_\xi$ exists. An explicit expression for
this generator will be derived in the next section.
Note that eq.~(\ref{hvf}) only fixes $\H_\xi$ up to terms that
have vanishing variation, i.e., terms that are defined entirely in terms of
the background structure, which in our case is the geometry of the boundary.

It is natural to fix this non-uniqueness by requiring that $\H_\xi(g_0)
= 0$ for all asymptotic symmetries in exact AdS.  Now, note that
given an exactly AdS metric $g_0$ we may take any asymptotic
symmetry $\eta$ to be represented by a Killing field of $g_0$.
As a result, the change $\delta_\eta {\cal H}_\xi$ may be
evaluated by taking $\delta g = {\pounds_\eta} g_0=0$ in
(\ref{hvf0}). Thus we find
\begin{equation}
\{\H_\eta, \H_\xi\}(g_0) = \delta_\eta \H_{\xi}(g_0) =0.
\end{equation}
As a result, taking $\H_\xi(g_0) =0$ sets $c(\xi, \eta)=0$ in
(\ref{pbracket}) and ensures that our generators satisfy the
symmetry algebra.

Furthermore, under the assumption that the symplectic structure $\sigma_\Sigma$ is independent of $\Sigma$, taking $\H_\xi(g_0) =0$ guarantees that $\H_\xi$ is
independent of the cut $C$.  This result is manifest when $\H_\xi$
is evaluated on $g_0$.  For a general asymptotically AdS-metric, we may
use the fact that (\ref{hvf0}) is independent of $\Sigma$ to establish that the failure (if any)
of $\H_\xi$ to be independent of the cut $C$ is given by an expression that has vanishing variation.
Assuming that any asymptotic AdS-geometry can be connected to $g_0$ by a path, it follows that
this expression in fact has to vanish.
We will establish this explicitly for the theory of interest at the end of section 3 below.

It is worth contrasting the case of $\Lambda < 0$ with asymptotically AdS boundary
conditions with the case $\Lambda = 0$ and asymptotically {\em flat} boundary
conditions.  In the latter case, it is found~\cite{hi, wz} that eq.~(\ref{consistency}) is
{\em not} satisfied and consequently, no ``absolutely conserved'' generator exists in that
case. This is directly related to the fact that, in the asymptotically flat case,
gravitational radiation can leak out to infinity.
In \cite{hi} it is
shown how to construct a canonical ``non-conserved'' generator
$\H_\xi[C]$ (equal to the Bondi energy-momentum and angular momentum) satisfying the
``balance law'' $\H_\xi[C_1] - \H_\xi[C_2] = \int_{\I_{12}} F_\xi$, where $F_\xi$ is a suitably defined flux-density\footnote{When $\xi^a$ is a time-translation, that flux is given by the square of a suitably defined ``news tensor''.}
through the portion $\I_{12}$ of scri bounded by $C_1$ and $C_2$.

\section{AdS charges}
\label{s3}

We now implement the strategy laid out in the previous section.
The upshot of our analysis will be that a conserved generator
$\H_\xi$ exists for every asymptotic symmetry
$\xi^a$ under our boundary conditions. It is given by
\ben
\label{hdef}
\H_\xi = \frac{-\ell}{8\pi G} \int_C \tilde E_{ab} \tilde u^a \xi^b \, d\tilde S.
\een
where $d\tilde S$ is the integration element on $C$ obtained from the unphysical metric,
where $\tilde u^a$ is the unit timelike
normal to $C$ within $\I$ (normalized with respect to the unphysical metric),
and where $\tilde E_{ab}$ is the normalized (leading order)
electric part of the unphysical Weyl tensor,
\ben
\label{Eabdef}
\tilde E_{ab} = \frac{1}{d-3} \Omega^{3-d} \tilde C_{acbd} \tn^c \tn^d \,.
\een
The vector field $\tn^a$ is defined as above by $\tn^a = \tnabla^a \Omega$, and
we remind the reader of our
convention that tensor indices on ``tilde'' quantities are raised and lowered with the unphysical
metric, while indices of ``untilde'' quantities are raised and lowered with the physical metric.
 It will be shown below that, despite the inverse powers of
$\Omega$, the quantity $\tilde E_{ab}$ is finite at $\I$ when the metric satisfies
our boundary conditions and Einstein's equation.

Expression~(\ref{hdef}) agrees with the expression
proposed previously by Ashtekar and Magnon~\cite{asht1} (in $d=4$) and by Ashtekar and Das~\cite{asht2} (for higher
dimensions). We emphasize, however, that our strategy leading to this expression is logically
rather different from that of Ashtekar et al. First, we {\em derive} this expression within a
Hamiltonian framework, whereas that expression was essentially guessed by Ashtekar et al.,
based on dimensional considerations,
on the fact that it reproduces the known expressions for energy-momentum and
angular momentum in the exactly known AdS-black hole spacetimes, and
on the fact that $\H_\xi$
turns out to be independent of the cross section $C$ (see below). Secondly, while Ashtekar et
al. essentially assume the finiteness of $\tilde E_{ab}$ as part of their boundary conditions, we in fact
derive the finiteness of $\tilde E_{ab}$.

As explained in the previous section, the fact that $\H_\xi$ is derived from a consistent Hamiltonian
framework together with the statement that $\H_\xi$ vanishes when evaluated on exact AdS space automatically implies that it is conserved, i.e., does not depend on the cross section
$C$. Actually, as pointed out by Ashtekar et al., this can also be seen explicitly. To see this,
one notes that, by definition, $\tilde E_{ab}$ is
trace-free and symmetric. It will be shown below that furthermore
\ben
\tilde D^a \tilde E_{ab} = 0 \quad {\rm on} \,\, \I,
\een
where $\tilde D_a$ is the intrinsic unphysical derivative operator on $\I$, i.e., the derivative
operator of $e^\omega(-dt^2 + d\sigma^2)$. Now the difference between
the hamiltonian charge for different cuts $C_1, C_2$ of $\I$ can be
written as
\ben
\label{balance}
\H_\xi[C_1] - \H_\xi[C_2] = - \frac{-\ell}{8 \pi G} \int_{\I_{12}} \tilde
D^a (\tilde E_{ab} \xi^b)
\, d\tilde s,
\een
using Stoke's theorem, where $d\tilde s$ is now the
$(d-1)$-dimensional integration element on $\I$ associated with the
unphysical metric. But it follows from the
properties of $\tilde E_{ab}$ that, on $\I$,
\ben
\tilde D^b (\tilde E_{ab} \xi^a) = \tilde E^{ab} \tilde D_{(a} \tilde
\xi_{b)} =
\frac{1}{d-1} \tilde E_a{}^a \tilde D_b \xi^b = 0,
\een
where we have used that an asymptotic symmetry restricts
to a conformal Killing field on $\I$. Hence,
the integrand of $\H_\xi$ is divergence free, and
we conclude that the integral does not depend on the choice
of the cut $C$.

Let us now derive the expression for $\H_\xi$, and the properties of $\tilde E_{ab}$.
As explained in the previous section, $\H_\xi$ is uniquely defined by eq.~(\ref{hvf})
specifying its variation $\delta \H_\xi$, and by the property that $\H_\xi = 0$ in
exact AdS space. The second property is obviously satisfied, because exact AdS
space has a vanishing Weyl tensor, and hence a vanishing $\tilde E_{ab}$.
In order to verify that our expression for $\H_\xi$ has the correct variation
postulated by eq.~(\ref{hvf}), we must analyze the consequences of Einstein's equations.
The reader not interested in the details of this analysis can jump directly to
lemma~3.1, where we summarize the results.

To analyze Einstein's equations, it is convenient to introduce the tensor field
\ben
\label{Sdef}
\tilde S_{ab} = \frac{2}{d-2} \tilde R_{ab} - \frac{1}{(d-1)(d-2)} \tilde R \tg_{ab},
\een
i.e., $\tilde S_{ab}$ is essentially the Ricci tensor of the unphysical metric $\tilde g_{ab}$.
In terms of this field, Einstein's equation is
\ben
\label{ee}
0 = \tilde S_{ab} + 2 \Omega^{-1} \tnabla_a \tn_b
- \Omega^{-2} \tg_{ab}(\tn^c \tn_c -\ell^{-2})\,.
\een
From now on, we set
\ben
\ell = 1 \,.
\een
Multiplying eq.~(\ref{ee}) by $\Omega^2$ and evaluating the result at $\I$, we see that
\ben
\label{n2=1}
\tn^c \tn_c \restriction \I = 1,
\een
i.e., $\tn^a$ is spacelike, unit, and normal to $\I$, consistent with
our assumption that $\I$ is a timelike boundary.
Now it is always possible to choose the conformal factor $\Omega$ so that
the unphysical metric takes the ``Gaussian normal form''
\ben
\tilde g_{ab}  = \tilde \nabla_a \Omega \tilde \nabla_b \Omega + \tilde h_{ab},
\een
where $\tilde h_{ab} \equiv \tilde h_{ab}(\Omega)$ is such that $\tilde h_{ab}(\Omega = 0)$
is equal to the metric of the Einstein static universe on $\I$, and such
that
\ben
\label{noshift}
\tilde h_{ab} \tnabla^b \Omega= 0, \quad \tg^{ab} \tnabla_a \Omega \tnabla_b \Omega   = 1
\een
in a full neighborhood of $\I$, (as usual, indices on tilda tensor
fields are raised and lowered with $\tilde g_{ab}$). In other words,
$\tilde h_{ab}$ is the induced metric on the surfaces $\I_\Omega$, the timelike
surfaces of constant $\Omega$. For example, for the metric of exact AdS space, the
choice of the conformal factor given by eq.~(\ref{unphysads}) satisfies eqs.~(\ref{noshift}), and the form
of $\tilde h_{ab}$ can be read off from expression~(\ref{unphysads}) as
\ben
\label{habexactads}
\tilde h_{ab} \, dx^a dx^b = - \left(1 + \frac{1}{4} \Omega^2 \right)^2 \, dt^2 +
\left( 1 - \frac{1}{4} \Omega^2 \right)^2 \, d\sigma^2 \quad {\rm in \,\, exact \,\, AdS}.
\een
For a general asymptotically AdS metric, a conformal factor satisfying eq.~(\ref{noshift})
may be found by first choosing an arbitrary $\Omega$, and then modifying this choice
if necessary as $\Omega \to  e^\omega \Omega$, where $\omega$ is to be determined.
One way to do this is by making a (formal) power series ansatz
$\omega = \sum \omega_i \Omega^i$, where ${\pounds_{\tn}} \omega_i = 0$. One
chooses $\omega_0$ so that the induced unphysical metric on $\I$ is the Einstein
static  universe, and $\omega_1$ so that $\tilde \nabla_a \tilde n_b = 0$ on $\I$.
Einstein's equations~(\ref{ee}) then immediately show that with these choices,
we have $\tilde n^c \tilde n_c = 1 + O(\Omega^2)$. It can be seen from this that the
remaining $\omega_2, \omega_3, \dots$ may then be chosen recursively
so that $\tilde n_c \tilde n^c = 1$ for the new choice of conformal factor, to all orders in $\Omega$.  Consequently, we can achieve that eq.~(\ref{noshift}) holds
to all orders in $\Omega$, and that the induced metric on $\I$ is the Einstein static universe.
For the arguments given below, it will not matter whether eq.~(\ref{noshift}) holds exactly,
or only to all orders in $\Omega$. For simplicity, we will assume that it holds exactly.

For a  conformal factor satisfying (\ref{noshift}), Einstein's equation~(\ref{ee}) simplifies to
\ben
\tilde S_{ab} = - 2 \Omega^{-1} \tilde \nabla_a \tn_b.
\een
We now use the standard technique of splitting this equation into
its components parallel and normal to $\I_\Omega$ (the surfaces of constant $\Omega$)
in a manner similar to the split performed in the ADM-formalism. If this is done,
then the following set of ``constraint'' and ``evolution'' equations
is obtained. The constraint equations are
\bena
-\tilde \R - \tilde \K_{ab} \tilde \K^{ab} + \tilde \K^2
+ 2(d-2) \Omega^{-1} \tilde \K
&=& 0 \,, \\
\tilde D^a \tilde \K_{ab} - \tilde D_b \tilde \K &=& 0 \,,
\eena
where $\tilde D_a$ is the derivative operator associated with $\tilde h_{ab}$,
 $\tilde \K_{ab} = - \tilde h_a{}^c \tilde h_b{}^d \tilde \nabla_c \tn_d$
is the extrinsic curvature of the surfaces $\I_\Omega$ (with respect to the unphysical
metric), and where $\tilde \R_a{}^b$ is the intrinsic Ricci tensor.
The evolution equations are\footnote{
The symbols ``$d/d\Omega$'' should be understood geometrically
as the Lie derivative along $\tn^a$.}
\bena
\label{evol1}
\frac{d}{d\Omega} \tilde \K_a{}^b &=& -\tilde \R_a{}^b + \tilde \K \tilde \K_a{}^b +
\Omega^{-1}(d-2) \tilde \K_a{}^b + \Omega^{-1} \tilde \K \delta_a{}^b, \\
\label{evol2}
\frac{d}{d\Omega} \tilde h_{ab} &=& -2 \tilde h_{bc} \tilde \K_a{}^c.
\eena
By assumption, $\I$ is a smooth boundary, implying that the fields
$\tilde h_{ab}, \tilde \K_{ab}$ must be smooth in a neighborhood of $\I$.
Consequently, multiplying the first evolution equation by
$\Omega$ and evaluating on $\I$, we immediately conclude that
\ben
\label{k0=0}
\tilde \K_{ab}  \restriction \I = 0 = \frac{d}{d\Omega} \tilde h_{ab} \restriction \I.
\een
To investigate more systematically the consequences implied by
eqs.~(\ref{evol1}, \ref{evol2}),
we express them in terms of the traceless part $\tilde p_a{}^b$
of $\tilde K_a{}^b$ and use the familiar technique
(see e.g. \cite{FG,KS6,Anderson}) of performing the Taylor expansions:
\ben
\tilde h_{ab} = \sum_{j=0}^\infty \Omega^j (\tilde h_{ab})_j, \quad
\tilde p_{a}{}^b = \sum_{j=0}^\infty \Omega^j (\tilde p_{a}{}^b)_j,
\een
where each tensor $(\tilde h_{ab})_j, \,(\tilde p_{a}{}^b)_j$ is independent
of $\Omega$ in the sense that the Lie-derivative along $\tilde n^a$ vanishes.
This yields the following recursion relations:
\bena
(d-2-j)(\tilde p_a{}^b)_j &=&
(\tilde \R_a{}^b)_{j-1} - \frac{1}{d-1}(\tilde \R)_{j-1} \delta_a{}^b
\nonumber \\  && {}\,
 - \sum_{m=0}^{j-1} (\tilde \K)_m (\tilde p_a{}^b)_{j-1-m} \,,
\label{recursionp}
\\
(2d-3-j)(\tilde \K)_j & =& (\tilde \R)_{j-1}
                       - \sum_{m=0}^{j-1} (\tilde \K)_m (\tilde \K)_{j-1-m} \,,
\label{recursionk}
\eena
as well as
\ben
j(\tilde h_{ab})_j
= -2\sum_{m=0}^{j-1}
  \Bigg[ (\tilde h_{bc})_m (\tilde p_a{}^c)_{j-1-m}
         + \frac{1}{d-1} (\tilde h_{ab})_m (\tilde \K)_{j-1-m} \Bigg] \,,
\label{recursionh}
\een
where we remind the reader that these equations hold for the
choice~(\ref{noshift}) of the conformal factor.
The ``initial conditions'' are, from eq.~(\ref{k0=0})
\ben
(\tilde p_a{}^b)_0 = (\tilde \K)_0 = 0 \,,
\een
and that $(\tilde h_{ab})_0 = -(dt)_a (dt)_b + \sigma_{ab}$ be the metric of the Einstein static universe.
The key point about these equations is that $(\tilde h_{ab})_j$ and
$(\tilde \K_a{}^b)_l$ are uniquely determined in terms of the initial
conditions for $j<d-1$ and $l<d-2$.
Therefore, they must be equal to the corresponding quantities
for exact AdS space, i.e., they are entirely ``kinematical''.
Thus, any quantity that depends only on $(\tilde h_{ab})_j$ and
$(\tilde \K_a{}^b)_l$ for $j$ and $l$ in this range,
must automatically be equal to
the corresponding quantity in pure AdS space. In particular, since
the Weyl tensor vanishes identically in pure AdS space, it follows that
\ben
(\tilde C^a{}_{bcd})_j = 0 \quad {\rm for} \,\, j < d-3
\een
in {\em any} asymptotically AdS spacetime satisfying the Einstein equations.
Furthermore, since $\tilde h_{ab}$ for exact AdS space is given by
equation~(\ref{habexactads}) it follows that
\ben
\tilde h_{ab} \, dx^a dx^b
= - \left(1 + \frac{1}{4} \Omega^2 \right)^2 \, dt^2
+ \left( 1 - \frac{1}{4} \Omega^2 \right)^2 \, d\sigma^2 + O(\Omega^{d-1})\,.
\label{h:confAdS}
\een
We will use these results later.

Let us now look at the recursion relation~(\ref{recursionp}) for $j=d-2$.
The left side of this equation is given by $0 \cdot (\tilde p_a{}^b)_{d-2}$,
so it does not yield any restriction on this coefficient\footnote{
Note that the the right side of this equation also vanishes, since
it vanishes in pure AdS, and since all the coefficients appearing
on the right side at this order are identical to the ones in pure AdS.
This is a non-trivial consistency check for our asymptotic conditions on
the metric.}.
The only restriction on this term comes from the constraint equation,
which fixes the divergence of $(\tilde p_a{}^b)_{d-2}$ with respect
to the derivative operator associated with the boundary metric.
It can be seen that, once a traceless, symmetric tensor
$(\tilde p_{a}{}^b)_{d-2}$ with the prescribed divergence is given, all tensors
$(\tilde p_a{}^b)_j, \,(\tilde h_{ab})_j$ are uniquely determined
for $j \ge d-1$ via the evolution and constraint equations.
Thus, this tensor carries the full information about the metric $g_{ab}$
which is not already supplied by the boundary conditions, i.e.,
the ``non-kinematical'' information. The tensor $(\tilde p_a{}^b)_{d-2}$ is
directly related to the leading order
electric part of the unphysical Weyl tensor, as we will now
show.

From the definition of the tensor field
$\tilde S_{ab}$, we
have
\ben
\label{RCS}
\tilde R_{abcd} = \tilde C_{abcd} + \tilde g_{a[c} \tilde S_{d]b}
- \tilde g_{b[c} \tilde S_{d]a} \,.
\een
Einstein's equation tells us that
$\tilde S_{ab} = 2 \Omega^{-1} \tilde K_{ab}$ for
a conformal factor satisfying (\ref{noshift}).
We substitute this relation into eq.~(\ref{RCS}), then contract
the resulting identity into $\tn^b \tn^d$ and project the
remaining free indices with $\tilde h_a{}^b$, yielding
\bena
\label{11}
 \tilde h^e{}_a\tilde h^f{}_c {\tilde R}_{ebfd}\tn^b\tn^d
 = \tilde h^e{}_a\tilde h^f{}_c {\tilde C}_{ebfd}\tn^b\tn^d  + \Omega^{-1}
\tilde K_{ac} \,.
\eena
Now, from the definition of the Riemann tensor and the extrinsic curvature
tensor, we have
\bena
\tilde h^e{}_a\tilde h^f{}_c {\tilde R}_{ebfd}\tn^b \tn^d
&=& \tilde h^e{}_a\tilde h^f{}_c
 \tn^b(\tilde \nabla_e \tilde \nabla_b - \tilde \nabla_b \tilde \nabla_e) \tn_f
\nonumber \\
&=& \pounds_\tn \tilde K_{ac} + \tilde K_{ab} \tilde K^b{}_c \,.
\eena
Substituting this into eq.~(\ref{11}) yields the equation
\ben
  \tilde C^a{}_{bcd} \tn^b \tn^d =
 \tilde h^{ea}\tilde h^f{}_c {\tilde C}_{ebfd}
 \tn^b \tn^d
 = - \tilde K^a{}_b \tilde K^b{}_c + {\pounds_{\tilde n}} \tilde K^a{}_c
  - \Omega^{-1} \tilde K^a{}_c \, .
\label{A-Key-equation-2}
\een
We can expand this equation in powers of $\Omega$, and remember
that the Lie derivative with respect to $\tilde n^a$ is identical
to $\partial_\Omega$ for conformal factors satisfying (\ref{noshift}).
We thereby obtain equations for the expansion coefficients.
At order $d-3$, we find the relation
\ben
\label{weylrecurs}
\frac{1}{d-3} (\tilde C_{acbd} \tn^c \tn^d)_{d-3}
= (\tilde K_{ab})_{d-2}  + \frac{1}{d-3} \sum_{m=0}^{d-3}
(\tilde K_{ac})_m(\tilde K_b{}^c)_{d-3-m} \,.
\een
Since, $\tilde h_{ab}$ is given by eq.~(\ref{habexactads}) for our choice of conformal factor in AdS space,
we know that $(\tilde h_{ab})_{d-1}$ vanishes in that case
(assuming $d \ge 6$). Consequently, $(\tilde K_{ab})_{d-2}$ also vanishes
in that case. Therefore, since the Weyl tensor vanishes in pure AdS,
the sum on the right side of eq.~(\ref{weylrecurs}) has to vanish in pure AdS.
However, the coefficients appearing in the sum are the same in any
asymptotically AdS spacetime when the conformal frame is chosen such that  boundary metric $(h_0)_{ab}$ given by the Einstein static universe. Thus,  the sum must in fact vanish
in all such cases. Consequently, we obtain the relation
\ben
\label{Cabcdge6}
\frac{1}{d-3} (\tilde C_{acbd} \tn^c \tn^d)_{d-3}
=  (\tilde K_{ab})_{d-2} \quad \mbox{when \,\,\,$d\ge 6$} \,
\een
for any asymptotically AdS spacetime in the appropriate conformal frame.

The remaining cases $d=4,5$ can be treated as follows:
In $d=4$, the sum on the right side of eq.~(\ref{weylrecurs}) reduces to
$2 (\tilde K_{ac})_1 (\tilde K^c{}_b)_0$, which vanishes by eq.~(\ref{k0=0}).
In $d=5$, the sum reduces to $\frac{1}{2}(\tilde K_{ac})_1 (\tilde K^c{}_b)_1$,
which can be expressed in terms of boundary data by the evolution equation.
A direct calculation using eqs.~(\ref{recursionp}, \ref{recursionk},
\ref{recursionh})
gives $\frac{1}{2}(\tilde K_{ac})_1 (\tilde K^c{}_b)_1
= \frac{1}{8} (\tilde h_{ab})_0$, therefore
\ben
\label{Cabcd5}
\frac{1}{2} (\tilde C_{acbd} \tn^c \tn^d)_{2} - \frac{1}{8} (h_{ab})_0
= (\tilde K_{ab})_{3}
\quad {\rm for} \,\,\, d=5.
\een
Equations~(\ref{Cabcd5}, \ref{Cabcdge6}) give the desired relation between the electric part of
the Weyl tensor at order $\Omega^{d-3}$, and $(\tilde K_{ab})_{d-2}$ (and therefore
also to $(\tilde p_a{}^b)_{d-2}$) for our choice of conformal factor.

Let us now look at recursion relation~(\ref{recursionh}) for $j=d-1$.
This can be written as
\ben
(\tilde h_{ab})_{d-1} = -\frac{2}{d-1} \,
(\tilde K_{ab})_{d-2} = - \frac{2}{d-1} (\tilde E_{ab})_0 \quad {\rm for} \,\, d=4, d\ge 6,
\een
where we have used eq.~(\ref{Cabcdge6}) and the definition of $\tilde E_{ab}$,
while
\ben
(\tilde h_{ab})_{4} = -\frac{1}{2} (\tilde E_{ab})_0 + \frac{1}{16} (\tilde h_{ab})_0, \quad {\rm for} \,\, d=5.
\een
Using eq.~(\ref{h:confAdS}), we find that the unphysical
line element can be written in the form
\bena
d \tilde s^2 &=& d\Omega^2 - \left(1 + \frac{1}{4} \Omega^2 \right)^2 \, dt^2
+ \left( 1 - \frac{1}{4} \Omega^2 \right)^2 \, d\sigma^2 \nonumber\\
&& - \frac{2}{d-1} \, \Omega^{d-1} \tilde E_{ij} \,
dx^i dx^j + O(\Omega^{d})\,  \quad {\rm for}\,\, d\ge 5,
\label{unphysmetric:678911}
\eena
while
\bena
d \tilde s^2 &=& d\Omega^2 - \left(1 + \frac{1}{2} \Omega^2 \right) \, dt^2
+ \left( 1 - \frac{1}{2} \Omega^2 \right) \, d\sigma^2 \nonumber\\
&& - \frac{2}{3} \, \Omega^{3} \tilde E_{ij} \,
dx^i dx^j + O(\Omega^{4})\,  \quad {\rm for}\,\, d=4 \, .
\label{unphysmetric:45}
\eena
In these expressions, $x^i$ are coordinates on $\I$, say $t$ and $d-2$ angles
parametrizing $S^{d-2}$, and $\Omega$ has been chosen so
that eq.~(\ref{noshift}) holds. This is the second key result of our analysis.

Let us finally derive that the leading order electric part of the Weyl tensor is divergence free.
This can be seen in different ways, for example by combining
eqs.~(\ref{Cabcdge6}, \ref{Cabcd5}) with the
constraint equation. Another way is to notice that, by combining Einstein's equation
and the contracted Bianchi-identities, we have
\ben
0 = \tilde \nabla^a( \Omega^{3-d} \tilde C_{abcd}),
\een
for any choice of the conformal factor. We now contract this equation into
$\tilde n^b \tn^c$ we use that $\tilde C_{abcd} = O(\Omega^{d-3})$,
and that
\ben
\tilde \nabla_a \tilde n_b = \frac{1}{d} (\tilde \nabla^c \tilde n_c) \tilde g_{ab} \quad
{\rm on} \,\, \I \, ,
\een
which follows from Einstein's equation. From this,
one immediately arrives at the relation $\tilde D^a \tilde E_{ab} = 0$ at $\I$ (for
any choice of the conformal factor).

Let us summarize what we have proved so far in a lemma.

\medskip
\noindent
\paragraph{Lemma 3.1:}
Let $(M, g_{ab})$ be an asymptotically AdS-spacetime satisfying Einstein's equation, with conformal completion $(\tilde M, \tilde g_{ab}, \Omega)$. Then the unphysical ($=$ physical) Weyl tensor
$\tilde C^a{}_{bcd}$ is of order $O(\Omega^{d-3})$, and the leading order electric part of the Weyl
tensor,  $\tilde E_{ab}$, given by eq.~(\ref{Eabdef}), satisfies $\tilde D^a \tilde E_{ab} = 0$ on $\I$, for any choice of the conformal factor. If the conformal factor is chosen as in eq.~(\ref{noshift}), then the unphysical metric can be expanded as
eqs.~(\ref{unphysmetric:45}, \ref{unphysmetric:678911})
near $\I$.

\medskip

We are now ready to prove the formula for $\H_\xi$. Consider a
metric $g_{ab}$ satisfying our asymptotic AdS condition.
We have seen that we may choose a conformal factor $\Omega$ such that
the unphysical metric $d\tilde s^2$ takes the form
eqs.~(\ref{unphysmetric:678911},\ref{unphysmetric:45}), and so
that the physical metric consequently takes the form $ds^2 =
\Omega^{-2} d\tilde s^2$. We may view
eqs.~(\ref{unphysmetric:678911},\ref{unphysmetric:45}) as a
``gauge condition'' on the metric, i.e., as picking a particular
representative in the equivalence class of metrics that are
diffeomorphic to $g_{ab}$. In this view, $\Omega$ is then a fixed
function on $\tilde M$ which is part of the background structure
specifying the asymptotic conditions. The (on-shell) metric
variations respecting this gauge choice (with $\Omega$ now
regarded as fixed) therefore take the form \ben \delta g_{ab} =
\gamma_{ab} + \pounds_\eta g_{ab} \een where the first piece
$\gamma_{ab}$ is a metric variation of the form
 \ben
\label{gammavar}
\gamma_{ab} = - \frac{2}{d-1} \, \Omega^{d-1} \delta \tilde E_{ab}
+ O(\Omega^d)\, ,
\een
(for a fixed choice of $\Omega$), and where the second piece is an infinitesimal diffeomorphism
generated by an arbitrary vector field $\eta^a$ respecting the gauge choice, i.e.,
a diffeo satisfying ${\pounds_\eta} ds_0^2 = O(\Omega^d)$, where $ds_0^2$ is the
line element of exact AdS-space~(\ref{ds00}). Thus,
\ben
\label{lieeta}
{\pounds_\eta} g_{ab} = - \frac{2}{d-1} \, \Omega^{d-1} {\pounds_\eta} \tilde E_{ab}
+ O(\Omega^d).
\een

Inserting these expressions into the definition of the symplectic form
$\omega(g, \delta_1 g, \delta_2 g)$,
we see that $\omega \restriction \I = 0$. Hence, the consistency condition~(\ref{consistency})
is satisfied, and we conclude by the general arguments given in the previous
section that $\H_\xi$ must exist.

Let us now determine the actual form of $\H_\xi$. For this, the
variations~(\ref{lieeta}, \ref{gammavar}) may be analyzed separately.
Let us calculate $\delta Q_\xi$ and $\xi \cdot \theta$
for the variation $\delta g_{ab} = \gamma_{ab}$. We can bring $Q_\xi$ in the form
\bena
(Q_\xi)_{a_1 \dots a_{d-2}}
= \frac{1}{8 \pi G} \Omega^{1-d} \tilde \epsilon_{a_1 \dots a_{d-2} bc}
\tn^b \xi^c
+ \frac{-1}{16 \pi G} \Omega^{2-d}
\tilde \epsilon_{a_1 \dots a_{d-2}bc} \tilde g^{be}
\tilde \nabla_e \xi^c \,.
\eena
We now take a variation of this expression and use
the relations $\delta \xi^a = 0$,
\bena
\delta \tilde \epsilon_{ab \dots c}
&=& -\frac{1}{d-1} \Omega^{d-1} \delta \tilde E_d{}^d \,
\tilde \epsilon_{ab \dots c} + O(\Omega^d) = O(\Omega^d) \,, \\
\delta \tn^a &=&
- \tilde g^{ac}\delta \tilde g_{cb} \tn^b
= +\frac{2}{d-1} \Omega^{d-1}
\delta \tilde E^{ab} \tn_b + O(\Omega^d) = O(\Omega^d) \,, \\
\delta (\tilde g^{be} \tilde \nabla_e \xi^c)
 &=& - 2\Omega^{d-2} \tilde n^{[b} \delta \tilde E^{c]}{}_d \xi^d
     - \Omega^{d-2} \delta \tilde E^{(bc)} n_d \xi^d
     + O(\Omega^{d-1}) \,.
\eena
This gives
\ben
\label{delqu}
(\delta Q_\xi)_{a_1 \dots a_{d-2}} = \frac{1}{8 \pi G}
\tilde \epsilon_{a_1 \dots a_{d-2}bc} \tn^b \delta \tilde E^c{}_d \xi^d
+ O(\Omega) \,.
\een
Using the relation
\ben
^{(d)}  {}\tilde \epsilon = \tilde n \wedge {}^{(d-1)} \tilde \epsilon = \tilde n \wedge \tilde u \wedge
{}^{(d-2)} \tilde \epsilon
\een
between the $d$-dimensional volume form, the induced $(d-1)$-dimensional
volume form on the boundary $\I$ and the $(d-2)$-dimensional volume form
on $C$, we can rewrite this
as
\ben
(\delta Q_\xi)_{a_1 \dots a_{d-2}} = \frac{1}{8 \pi G} \delta \Bigg[
{}^{(d-2)} \tilde \epsilon_{a_1 \dots a_{d-2}}
\, \tilde E^c{}_d \tilde u_c \xi^d \Bigg] \,, \,\,\, \mbox{on $\I$} \,.
\een
A similar calculation with the fact $g^{ab}\delta g_{bc}=O(\Omega^{d-1})$,
$\delta g^c{}_c=O(\Omega^d)$ gives
$\theta(g, \delta g = \gamma) \restriction \I = 0$.
Thus, eq.~(\ref{hdef}) gives
\ben
\delta \H_\xi = \int_C \delta Q_\xi = \frac{-1}{8 \pi G} \,\, \delta \int_C
\tilde E_{ab} \tilde u^b \xi^a \, d \tilde S.
\een
which is the defining relation for $\H_\xi$ (if $\ell$ is restored).

Since we have established that the integral on the right-hand side is conserved, it is unchanged by any diffeomorphism which acts on both the metric $g_{ab}$ and the asymptotic symmetry $\xi^a$.  As a result, the variation of this integral under $\delta g_{ab} = \pounds_\eta g_{ab}$ (while holding $\xi^a$ fixed) is given by replacing $\xi^a$ with $[\eta, \xi]^a$.  On the other hand, it can be verified directly from eq.~(\ref{lieeta})
and the definition of the symplectic structure~(\ref{symplectic})
that $\H_{[\xi, \eta]} = \sigma_\Sigma(g; \pounds_\xi g, \pounds_\eta g)$.
Thus equation (\ref{hdef}) does indeed satisfy the variational condition
(\ref{hvf0}) for such variations.  As a result,
we have shown that $\H_\xi$ given by eq.~(\ref{hdef})
is the correct expression for the Hamiltonian generator.

\section{Comparison to other definitions of conserved quantities}

As explained in the previous subsection, our definition of the conserved quantities $\H_\xi$
associated with asymptotic symmetries $\xi^a$
agrees with the one proposed by Ashtekar et al. However, there also exist other definitions of
conserved charges associated with asymptotically AdS spacetimes in the literature.
In the following subsections, we investigate the relation between some of those charges
and our definition above.

\subsection{The counterterm subtraction method}

In the counterterm subtraction method~\cite{skenderis,kraus,KS2,KS3,KS4,KS5,KS6,KS7,KS8,KS9}, ``charges'' $\Q_\xi$ associated with
asymptotic symmetries $\xi^a$ are constructed from
an ``effective energy momentum tensor'', $\tau_{ab}$, which is obtained by varying an
auxiliary ``effective boundary Lagrangian'' (not to be confused with
the Einstein Lagrangian given in
eq.~(\ref{einsteinhilbert})).
They are defined by
\ben
\label{qcounterdef}
\Q_\xi = \lim_{C \to \I} \int_C \tau_{ab} \xi^a \hat u^b \, dS,
\een
where $C$ is a sequence of cross sections that is taken to $\I$ within a Cauchy surface
$\Sigma$, and where $\hat u^a$ is the unit normal to that surface.
The charges $\Q_\xi$ are quite different, both in appeareance (see below) and conceptually, from
the Hamiltonian charges $\H_\xi$, so it is natural to investigate the relation between the two.
A first
step in this direction was taken by Ashtekar and Das~\cite{asht2}, who derived an explicit expression for
the difference between $\H_\xi$ and $\Q_\xi$ in terms of the extrinsic curvature of $\I$ in the
ambient manifold $\tilde M$. An explicit evaluation of this expression for pure AdS space shows that
$\H_\xi$ and $\Q_\xi$ differ, as also remarked in~\cite{kraus, asht2}. However,
the analysis of~\cite{asht2} left open the question whether the difference between $\Q_\xi$ and
$\H_\xi$ would in general be depend on the particular asymptotic AdS spacetime under consideration,
or whether that difference would only consist of a constant offset.   In addition, one may ask what
algebra the charges $\Q_\xi$ generate under Poisson bracket. We will now address these
issues.

The actual form of the effective boundary Lagrangian, and of $\tau_{ab}$, depends on the number of
spacetime dimensions, and becomes increasingly complicated for increasing $d$. To keep our discussion as simple as possible, and because of the relevance of that case to the AdS/CFT-correspondence, we restrict attention to the case $d=5$ (as is also done
in~\cite{asht2}). However, our arguments could, in principle, be extended to higher dimensions and one would expect a similar result.  A simple independent argument addressing the algebra generated by the ${\cal Q}_\xi$ in all dimensions will appear in \cite{peierls}.

Let $(M, g_{ab})$ be an asymptotically AdS spacetime in the sense of our definition, satisfying the
Einstein equations. Let $(\tilde M, \tilde g_{ab}, \Omega)$ be its conformal completion. We
denote by $\I_\Omega$ the surfaces of constant $\Omega$, so that, when $\Omega$ is small,
the $\I_\Omega$ are timelike surfaces. Let $\hat \eta^a$ be the inward
directed unit normal to $\I_\Omega$,
let $h_{ab}$ the intrinsic metric on $\I_\Omega$ defined by
\ben
h_{ab} \equiv -\hat \eta_a \hat \eta_b + g_{ab},
\een
let $K_{ab} = -h_a{}^c h_b{}^d \nabla_c \hat \eta_d$
be the extrinsic curvature, and let $\R_{ab}$
the Ricci tensor of $h_{ab}$ on $\I_\Omega$.

The total action for $5$-dimensions is given by
the Einstein-Hilbert~(\ref{lagrangian:Einstein}) plus
the boundary lagrangian
\ben
 \frac{1}{8 \pi G}\, \sqrt{-h} \, K d^4 x
- \frac{1}{8 \pi G} \, \sqrt{-h} \, (3 + \frac{1}{4} \R) \, d^4 x \,,
\een
where the first term corresponds to the familiar
``Gibbons-Hawking'' boundary term that may be added to the action
of general relativity, while the second and third term are quantities
that are constructed entirely out of the intrinsic geometry of $\I_\Omega$.
The variation of the total action with respect to the metric $h_{ab}$
provides the effective stress energy tensor
\bena
\tau_{ab}
&=&\frac{1}{8 \pi G}
\left[ \frac{1}{2}\left( \R_{ab} - \frac{1}{2} \R h_{ab} \right)
- 3 h_{ab} - K_{ab} + K h_{ab} \right].
\eena
Let us introduce
\ben
\tilde f = \Omega^{-2}(1- \tilde \eta^a \tilde \nabla_a \Omega),
\een
where $\tilde \eta^a$ is the inward unit normal to $\I_\Omega$
with respect to the metric $\tilde g_{ab}$, and let us express
the Gauss-Codacci equation
$
 C_{acbd} \hat \eta^c \hat \eta^d
= -{\cal R}_{ab} + K K_{ab}-K_{ac} K^c{}_b - (d-2)h_{ab}
$
in terms of the corresponding unphysical quantities by
using Einstein's equations. Then, using those expressions,
one finds
\bena
\label{tab-eab}
\tau_{ab} - \frac{-1}{16 \pi G} C_{acbd}
\hat \eta^c \hat \eta^d &=&
\frac{1}{16 \pi G} \Bigg[
\tilde K \tilde K_{ab} - \tilde K_a{}^c \tilde K_{cb} + \frac{1}{2}(\tilde K_{cd} \tilde K^{cd}
- \tilde K^2) \tilde h_{ab} \nonumber \\
&-& 2\Omega \tilde f(\tilde K_{ab} - \tilde K \tilde h_{ab}) - 3\Omega^2 \tilde f^2 \tilde h_{ab}
\Bigg]\\
&\equiv& \Omega^2 \tilde \Delta_{ab},
\eena
where we remind the reader that tilda quantities refer to the unphysical spacetime, i.e.,
$\tilde h_{ab} = \Omega^2 h_{ab}$ and
$\tilde K_{ab} = -\tilde h^c{}_a \tilde h^d{}_b \tilde \nabla_c
\tilde \eta_d$.
From eqs.~(\ref{k0=0}, \ref{n2=1}) we know
that $\tilde K_{ab} \restriction \I = 0 = \tilde f \restriction \I$,
implying that $\tilde \Delta_{ab}$ is finite at $\I$.
The difference between the counterterm
charge and the Hamiltonian charge can immediately
be found by integrating eq.~(\ref{tab-eab}) over $C$. It is given by
\ben
\label{qhdiff}
\Q_\xi - \H_\xi = \int_C \tilde \Delta_{ab} \tilde u^b \xi^a  \, d\tilde S,
\een
where $\tilde u^a$ is the unit normal (normalized with respect to $\tilde g_{ab}$) to $C$
within $\I$. This is the expression obtained by Ashtekar and Das~\cite{asht2}.

To understand better the structure
of the integrand $\tilde \Delta_{ab}$, one needs to further use Einstein's equations.
For this, we note that the unphysical metric can be written as
\ben
\tilde g_{ab} = (1-\Omega^2 \tilde f)^{-2} \tilde \nabla_a \Omega
\tilde \nabla_b \Omega + \tilde h_{ab},
\een
where $\tn^a = \tilde \nabla^a \Omega$, and $\tilde h_{ab}  \tn^b = 0$.
Using expansion techniques similar to those in the previous section,
one can infer from Einstein's equations that
\ben
\label{kabrab}
\Omega^{-1} \tilde K_{ab} = \frac{1}{2} \left( \tilde \R_{ab} - \frac{1}{6}
\tilde \R \tilde h_{ab} \right) + \tilde f \tilde h_{ab} \quad {\rm
on} \,\, \I.
\een
Substituting this back into eq.~(\ref{tab-eab}), one gets
\bena
\tilde \Delta_{ab} &=&
\frac{1}{64 \pi G}
 \Bigg(
       \frac{2}{3} \tilde \R \tilde \R_{ab}
       - \frac{1}{4} \tilde \R^2 \tilde h_{ab}
       - \tilde \R_{ac} \tilde \R^c{}_b
       + \frac{1}{2} \tilde \R_{mn} \tilde \R^{mn} \tilde h_{ab}
 \Bigg)
\eena
on $\I$, where we note that the terms involving $\tilde f$ have cancelled.
and where $\tilde {\cal G}_{ab} $ denotes the Einstein tensor on $\I$.
Therefore we find
\bena
\label{Zdef}
&&{\mathcal Z}_\xi \equiv \Q_\xi - \H_\xi \nonumber = \\
&&
\frac{1}{64 \pi G} \int_C \Bigg(
\frac{2}{3} \tilde \R \tilde \R_{ab} - \frac{1}{4} \tilde \R^2 \tilde h_{ab} - \tilde \R_{ac} \tilde \R^c{}_b +\frac{1}{2} \tilde \R_{mn} \tilde \R^{mn} \tilde h_{ab}
 \Bigg) \xi^a \tilde u^b \, d \tilde S \, ,
\eena
where we note that, so far, we have nowhere assumed that $\tilde h_{ab}$ is
the metric of the Einstein static universe\footnote{ 
Note that the trace of the integrand
of the right side corresponds to the trace of $\tau_{ab}$ and is given by
$\tau_a{}^a = \frac{1}{8 \pi G}(\frac{1}{8} \tilde \R_{ab} \tilde \R^{ab}
- \frac{1}{24} \tilde \R^2) $.
This corresponds to the ``trace-anomaly'' found in~\cite{kraus,skenderis}
in the context of the AdS-CFT correspondence.}.
Thus, the charge determined by the counterterm
subtraction method disagrees with $\H_\xi$. However, the difference between the two is just given
by a constant offset, which is determined in terms of the boundary metric $\tilde h_{ab}$ and its curvature
tensor, and which is hence independent of the actual asymptotically AdS solution. It can therefore be
evaluated in {\em any} asymptotically AdS solution with a given boundary metric $\tilde h_{ab}$, in
particular in pure AdS space.

For this reason, the difference ${\mathcal Z}_\xi$
has vanishing Poisson bracket with any observable and, in
particular, with the generators $\H_\xi$. Since the generators
$\H_\xi$ satisfy the algebra eq.~(\ref{pbracket}) under Poisson brackets,
the charges $\Q_\xi$ satisfy the algebra
\ben
\{ \Q_\xi, \Q_\eta \} = \Q_{[\xi, \eta]} - {\mathcal Z}_{[\xi, \eta]},
\een
i.e., a trivial central extension of the $O(d-1, 2)$ algebra.
In addition, we may note that
the counter-term energy is consistent with the
covariant phase space methods of \cite{wz}, which controls only variations
of the Hamiltonian on the space of solutions.

Let us now use the fact that, on $\I$, we may take $\tilde h_{ab} $ to be
the metric of the Einstein static universe,
$\tilde h_{ab} = -t_a t_b + \sigma_{ab}$ (with
$t^a = (\partial / \partial t)^a$).
This metric has the Ricci tensor
\ben
\tilde \R_{ab} = 2\sigma_{ab}.
\een
Inserting this into eq.~(\ref{Zdef}), we get
\ben
\label{Zdef1}
\Q_\xi - \H_\xi = \frac{1}{16 \pi G} \int_C \left(\frac{3}{4} t_a t_b +
  \frac{1}{4} \sigma_{ab} \right) \xi^a \tilde u^b \, d \tilde S.
\een
Choosing the symmetry to be a time translation, $\xi^a = t^a$, gives
\ben
\Q_t - \H_t = \frac{3A_3}{64 \pi G}
\een
where $A_3$ is the area of the unit 3-sphere. In particular, while $\H_t$ vanishes in
exact AdS space, the charge $\Q_t$ does not vanish and is given by the right side of
the above equation.  In the context of the AdS/CFT correspondence,
the above value of $\Q_t$ in pure AdS space is interpreted as the Casimir energy of the
CFT. However, our result implies the stronger statement that, in the above conformal frame, $\H_t -
\Q_t$ is given by expression~(\ref{Zdef1}) in {\em any} asymptotically AdS spacetime.

We finally note that $\Q_\xi$ is conserved for any asymptotic
symmetry $\xi^a$, in any asymptotic AdS spacetime, in the sense that it does not depend on the
cross section $C$ chosen to calculate $\Q_\xi$. This follows from the fact that $\H_\xi$ has
this property, and the fact that the integrand
$\frac{3}{4} t_a t_b + \frac{1}{4} \sigma_{ab}$ on the right side of
eq.~(\ref{Zdef}) is covariantly conserved on $\I$, and has a vanishing
trace\footnote{
Note that there is a claim to the opposite on p.13 in~\cite{asht2};
we suspect a calculation error in this reference.}.

\subsection{The Henneaux-Teitelboim definition}
\label{htsubsec}

Henneaux and Teitelboim~\cite{ht} consider asymptotically AdS boundary
conditions on the metric specified by
demanding that there exists a coordinate system
$x^\mu = (t, r, \theta^i)$
near infinity (with $\theta^i$ coordinates on $S^{d-2}$)
such that the line element $ds^2$ under consideration can be written as
\ben
ds^2 = ds^2_0 + \sum_{\mu, \nu} \gamma_{\mu \nu} \, dx^\mu dx^\mu
\label{htform}
\een
where $ds_0^2$ is the line element of exact AdS space given by eq.~(\ref{pureads}),
and where it is demanded that
\bena
\label{ht1}
\gamma_{tt} &=& O(r^{-d+3}), \\
\gamma_{rr} &=& O(r^{-d-1}), \\
\gamma_{tr} &=& O(r^{-d}), \\
\gamma_{r \theta^i} &=& O(r^{-d}), \\
\gamma_{t \theta^i} &=& O(r^{-d+3}), \\
\label{ht5}
\gamma_{\theta^i \theta^j} &=& O(r^{-d+3}).
\eena
If $\xi^a$ is a Killing vector field of exact AdS-space,
then the integrals in
\bena
\label{htcharge}
\Q_\xi &=& \int_\Sigma {\mathcal C}^a \xi_a \nonumber\\
&+&\lim_{C \to \I} \frac{1}{16 \pi G} \int_C G_a{}^{bdc}\left[ \xi^e \hat u_e D_b \gamma_{cd} -
\gamma_{cd} D_b (\xi^e \hat u_e) \right] \hat \eta^a \, dS \nonumber\\
&+& \lim_{C \to \I}
\frac{1}{4\pi G} \int_C (\kappa_{ab} - \kappa q_{ab}) \xi^a \hat \eta^b \, dS
\eena
converge as $C$ tends to a cross section at $\I$ within a spatial slice $\Sigma$.
Here, ${\mathcal C}^a$ are the constraints (viewed as $(d-1)$ forms),
$\hat u^a$ is the unit normal to $\Sigma$, $\hat \eta^a$ is the unit normal\footnote{In other
words, $\hat u^{[a} \hat \eta^{b]}$ is the binormal to $C$.}
to $C$ within $\Sigma$,  $q_{ab}= g_{ab} + \hat u_a \hat u_b$
is the induced metric on $\Sigma$, $D_a$ is the associated spatial
derivative operator, and $\kappa_{ab} =
-q_a{}^c q_b{}^d \nabla_c \hat u_d$ is the extrinsic curvature. Moreover,
\ben
G_{abcd} = \frac{1}{2} (q_{ac} q_{bd} + q_{ad} q_{bc} - 2q_{ab}q_{cd}).
\een
has been defined. Henneaux and Teitelboim take $\Q_\xi$ as the definition of
the conserved quantity associated with $\xi^a$ for a
spacetime satisfying the asymptotic conditions eqs.~(\ref{ht1}---\ref{ht5}). These conditions are,
in turn, motivated by the fact (i) that they hold in the familiar examples of black hole spacetimes
in the presence of a cosmological constant, (ii) that they are preserved when acting with
a diffeomorphism $\psi$ that is an exact symmetry of the pure AdS background,
and (iii) that they imply finiteness
of the charges $\Q_\xi$, and these can be shown to form a representation of
the asymptotic symmetry algebra under Poisson brackets.

The asymptotic conditions eqs.~(\ref{ht1}---\ref{ht5})
(including the precise notion of an asymptotic symmetry),
and the expression $\Q_\xi$ for  the charges are different
in appearance from our asymptotic conditions and the charges $\H_\xi$, but we will now show that
they are, in fact equivalent. Starting with the boundary conditions, assume that $(M, g_{ab})$
satisfies the asymptotic conditions proposed by Henneaux and Teitelboim. Then, defining e.g.
$\Omega = 1/r$, we see that the metric also satisfies our boundary conditions. Conversely,
assume that our boundary conditions hold and that Einstein's equation is satisfied. The conformal factor
$\Omega$ may then be chosen so that the unphysical metric $\tilde g_{ab} =
\Omega^2 g_{ab}$ has the form given in eqs.~(\ref{unphysmetric:45},
\ref{unphysmetric:678911}).
We choose coordinates $x^\mu$ as follows:
We define a coordinate $r$ by
\ben
r = \frac{1}{2}(\Omega^{-1} - \Omega),
\een
and on $\I$.   We choose coordinates $(t, \theta^\alpha)$ such that, on $\I$, the line element of the metric
$\tilde h_{ab}$
takes the form $-dt^2 + \sum \sigma_{ij} d\theta^i d\theta^j$, where $\sigma_{ij}$
are the coordinate components of the round metric on $S^{d-2}$.
A point $x$ in a neighborhood
of $\I$ may then be assigned coordinates $x^\mu = (t, r, \theta^i)$ in an arbitrary smooth manner.
Substituting $r$ in terms of $\Omega$ in eqs.~(\ref{unphysmetric:45},
\ref{unphysmetric:678911}) then
immediately implies that $ds^2 = \Omega^{-2} d\tilde s^2$ can be
written as $ds^2_0 + \sum \gamma_{\mu\nu} \, dx^\mu dx^\nu$, where the coordinate
components $\gamma_{\mu\nu}$ have the fall off given in eqs.~(\ref{ht1}---\ref{ht5}).
Thus, our boundary conditions and the ones considered
by Henneaux and Teitelboim are equivalent.

Let us compare next the respective notions of asymptotic symmetry. Let $g_{ab}$ be an
asymptotically AdS metric, written in the form~(\ref{htform})
for some choice of coordinates $x^\mu$. Henneaux and Teitelboim consider
vector fields $\xi^a$ that are exact symmetries of the underlying background AdS space chosen, implying that $\pounds_\xi g_{ab} = \pounds_\xi \gamma_{ab}$. The components $\pounds_\xi \gamma_{\mu\nu}$ can be checked to satisfy the fall off conditions given in eqs.~(\ref{ht1}---\ref{ht5}). Therefore,
the 1-parameter group $\psi_t$ of diffeos generated by $\xi^a$ has the property that if $g_{ab}$
is asymptotically AdS, then so is
$\psi_t^* g_{ab}$. Consequently, $\xi^a$ is an
asymptotic symmetry in our sense. Conversely, let $\xi^a$ be a vector field on $M$ that is an asymptotic
symmetry in our sense, let $g_{ab}$ the metric of exact AdS space, and let $x^\mu$ be
coordinates such that $ds_0^2 = \sum g_{\mu\nu} \, dx^\mu dx^\nu$ takes the form~(\ref{pureads}).
Now it is easy to see that, in a neighborhood of $\I$, one can find a diffeomorphism $\phi$ leaving $\I$ invariant such that $\xi^a$ is a Killing field for $\phi^*g_{ab}$. Now let $x^{\prime \mu}$ be
the coordinates related to $x^\mu$ via $\phi$. It follows that $\xi^a$ is a Killing field of
$\sum (g_{\mu\nu} \circ \phi) \, dx^{\prime \mu} dx^{\prime \nu}$,
which has the form~(\ref{pureads}).
Consequently, $\xi^a$ is an asymptotic symmetry in the sense of Henneaux and Teitelboim.

Let us finally compare the charges $\Q_\xi$ to our charges
$\H_\xi$. The charges $\H_\xi$ are uniquely determined by
the requirement that they satisfy eq.~(\ref{hvf}) and that
$\H_\xi=0$ in exact AdS space~\cite{wz}. However, the charges
$\Q_\xi$ of Henneaux and Teitelboim are constructed precisely
so as to satisfy these conditions as well~\cite{ht}. Hence they must
agree with our charges. Note that, since the definition of
$\H_\xi$ is manifestly ``background independent'', this
result implies that the same is true for the charges $\Q_\xi$.
even though the latter are not manifestly background independent,
due to the dependence of $\gamma_{ab}$ on the choice
of coordinates $x^\mu$ implicit in eq.~(\ref{htform}).

\subsection{The spinor definition}

Another way to define conserved charges associated with energy and momentum is
provided by the spinor method~\cite{w,warner,gary}. In this approach,
one makes use of the fact that the asymptotic symmetry group has a
spinor representation. Consider an asymptotically AdS spacetime $(M, g_{ab})$, and assume that
a spinor bundle, $S$, and corresponding curved space gamma matrices
$\gamma_a$ satisfying $\gamma_{(a} \gamma_{b)} = g_{ab}$ can be defined\footnote{Here one
needs to assume that $(M, g_{ab})$ admits a spinor bundle, which imposes some
mild restrictions on the topology of $M$.}. Given an asymptotic symmetry represented by
a future directed, timelike or null vector field $\xi^a$, one can find an auxiliary spinor field
$\psi$ such that $\xi^a = \overline{\psi} \gamma^a \psi$, up to terms that vanish at infinity.
One then considers the ``Nester 2-form''
\ben
B_{ab} = \frac{1}{8\pi G}
\left[\overline{\psi} \gamma_{[a} \gamma_b \gamma_{c]} \widehat \nabla^c \psi
- \overline{(\widehat \nabla^c \psi)} \gamma_{[a} \gamma_b \gamma_{c]} \psi\right],
\een
where overline denotes the Dirac conjugate of a spinor field,
and where $\widehat \nabla_a$ is the operator defined by
\ben
\widehat \nabla_a \psi = \nabla_a \psi - \frac{1}{2\ell} \gamma_a \psi
\een
in terms of the covariant derivative operator on spinor fields. (From now on we set $\ell = 1$.)
The spinor charge is defined
by\footnote{In differential forms notation, $\Q_\xi = \int_C {} \, *B$.}
\ben
\label{spinorcharge}
\Q_\xi = \lim_{C \to \I} \int_C B_{ab} \hat u^b \hat \eta^a \, dS,
\een
where $\hat u^a$ is the unit normal to the Cauchy surface $\Sigma$ in which the $(d-2)$ surfaces
$C$ are embedded and where $\hat \eta^a$ is the normal to $C$ within $\Sigma$. In order for
the limit to exist, the spinor field $\psi$ must be chosen in such a
way that $\widehat \nabla_a \psi$ vanishes sufficiently fast at
infinity.
Finally, one may prove that $\psi$ can, in addition, be chosen to satisfy the ``Witten condition'',
$q^{ab} \gamma_a \widehat \nabla_b \psi = 0$ on $\Sigma$ (with $q_{ab}$
given by $g_{ab} = -\hat u_a \hat u_b + q_{ab}$). If this condition is imposed\footnote{If $\Sigma$
has ``inner boundaries'' corresponding to horizon cross sections of black holes, then suitable
boundary conditions need to be imposed upon $\psi$ at those inner boundaries in addition
to the conditions upon the asymptotic behavior of $\psi$ at infinity.  Those
may be taken to be of the form
$\epsilon^{ab} \gamma_a \gamma_b \psi \restriction {\mathcal B} = 0$,
where $\epsilon^{ab}$ is the binormal to the horizon cross section
$\mathcal B$, see \cite{gary}.} then
$\Q_\xi$ can be brought into a form which is manifestly non-negative, and which vanishes
if and only if the spacetime under consideration is pure AdS. Therefore, since
the future directed, timelike or null asymptotic symmetries $\xi^a$ correspond to energy-momentum,
it follows that the charges $\Q_\xi$ give positive energy-momentum for any asymptotically
AdS spacetime, and it shows that the only spacetime with zero energy is AdS-space itself.

\medskip

It was shown in~\cite{davis} that the charge $\Q_\xi$ agrees with the Ashtekar-Magnon
definition, and hence with $\H_\xi$ in $d=4$ dimensions. This establishes in particular that
$\H_\xi$ is a non-negative quantity whenever $\xi^a$ is a
future pointing timelike or null asymptotic symmetry, and that $\H_\xi = 0$
implies that the spacetime is (isometric to) pure AdS. We will now show that $\H_\xi = \Q_\xi$
also in higher dimensions, thereby showing that $\H_\xi$ yields an energy
that is positive.

Consider first the case when the metric $g_{ab}$
is that of exact AdS space. Then it can be shown that there exist $d$ linearly independent
spinor fields $\alpha$ satisfying the ``Killing spinor equation''
\ben
\widehat \nabla_a \alpha = 0 \quad ({\rm exact \,\, AdS}).
\een
The vector field $\xi^a = \bar \alpha \gamma^a \alpha$ can then be seen to be
a (necessarily timelike or null) Killing vector field in AdS. The spinor
charge $\Q_\xi$ by definition vanishes for these symmetries, i.e., the energy-momentum
of pure AdS computed with the spinor charge vanishes, in agreement with the charges $\H_\xi$.

In order to investigate the relation between $\Q_\xi$ and $\H_\xi$ in general
asymptotically AdS spacetimes, we pass to a conformal completion $(\tilde M, \tilde g_{ab}, \Omega)$
with corresponding spinor bundle $\tilde S$ and gamma matrices $\tilde \gamma_{(a} \tilde \gamma_{b)}
= \tilde g_{ab}$. Spinor quantities (i.e., sections) in
$\tilde S$ or its tensor products can naturally be identified with sections in $S$ by putting
$\tilde \psi = \Omega^{1/2} \psi$ and $\tilde \gamma_a = \Omega \gamma_a$. With these
identifications, it then follows that
\ben
\overline{\tilde \psi} \tilde \gamma^a \tilde \psi = \overline \psi \gamma^a \psi,
\een
and it follows that $\tilde \gamma_a$ is smooth at $\I$. The relation between
the physical and unphysical derivative operators on spinors is given by
\ben
\nabla_a \psi = \tilde \nabla_a \psi + \frac{1}{2} \Omega^{-1}(\tilde \gamma_a \tilde \gamma_b \tn^b
+ \tn_a) \, \psi.
\een
As shown above, it is always possible to write the metric in the
form~(\ref{unphysmetric:678911}, \ref{unphysmetric:45})
when Einstein's equations are satisfied. The first line
in these expressions is the metric $d \tilde s_0^2 = \Omega^2 ds_0^2$ of
pure AdS space multiplied by a conformal factor.
We choose $\psi = \alpha$, where $\alpha$ is a Killing spinor of pure AdS.
Then $\tilde \psi = \Omega^{1/2} \psi$ is smooth at $\I$, and
using the fact that $\psi$ is a Killing spinor for $ds_0^2$,
it can then be verified that
\ben
\label{hatnablapsi}
\widehat \nabla_a \psi = -\frac{1}{2}
\Omega^{d-5/2} \tilde \gamma_c \tilde \gamma_d \tn^{[c} \tilde E^{d]}{}_a \tilde \psi
-\frac{1}{2(d-1)}\Omega^{d-5/2} \tilde \gamma_c \tilde E^c{}_a (\tilde \gamma_b \tilde n^b - 1) \tilde \psi + O(\Omega^{d-3/2}).
\een
It also
follows from the fact that $\psi$ is a Killing spinor for the pure AdS metric, that
\ben
\label{spinorid}
\tn^a \tilde \gamma_a \tilde \psi \restriction \I \, \, =  \tilde \psi \restriction \I.
\een
Substituting this expression into eq.~(\ref{hatnablapsi}), we get
\ben
\widehat \nabla_a \psi = \frac{1}{2} \Omega^{d-5/2} \tilde \gamma_c \tilde E^c{}_a \tilde \psi
+ O(\Omega^{d-3/2}).
\een
We now substitute this into the definition of $B_{ab}$, we contract the result with
$\hat \eta^{a} \hat u^{b}$, we use
that $\Omega \tilde u^a = \hat u^a$ and $\Omega \tilde n^a = \hat \eta^a$ on $\I$, and
we use the gamma-matrix identity
\ben
\tilde \gamma^{[b} \tilde \gamma^a \tilde \gamma^{d]}= \tilde \gamma^{[b} \tilde \gamma^{a]} \tilde \gamma^d + 2\tg^{d[b} \tilde \gamma^{a]} .
\een
If this is done,  one finds
\ben
B_{ab} \hat u^a \hat \eta^b = -\frac{1}{8\pi G} \Omega^{d-2} (\overline{\tilde \psi} \tilde \gamma^c
\tilde \psi) \,\, \tilde E_{cd} \tilde u^d + O(\Omega^{d-1}).
\een
We finally integrate this equation over $C$ and use that $dS = \Omega^{2-d} d\tilde S$,
as well as $\xi^a = \overline{ \tilde \psi } \tilde \gamma^a \tilde \psi$. This gives
\ben
\Q_\xi = \lim_{C \to \I} \int_C B_{ab} \hat u^a \hat \eta^b \, dS = \frac{-1}{8 \pi\, G} \int_C
\tilde E_{ab} \tilde u^b \xi^a  \, d \tilde S = \H_\xi,
\een
proving the equivalence of $\Q_\xi$ and $\H_\xi$ in $d$ dimensions.

\section{Inclusion of matter fields}

We now apply the formalism developed in sections~2 and~3 to derive the explicit formula for $\H_\xi$ in the case where additional (bosonic) matter fields are present.  The matter fields of most interest are those which appear in the supergravity theories relevant to various AdS/CFT conjectures; i.e., to scalars and anti-symmetric tensor fields.    An overview of the results is presented in section \ref{overview}, followed by a detailed treatment of the conserved energy in the presence of such fields using the covariant phase space approach (section \ref{scalars}), the counter-term subtraction approach (section \ref{cts}), and the spinor approach (section \ref{spinorchargesec}).


\subsection{Overview of results with matter fields}
\label{overview}

The effect of matter fields on $\H_\xi$ is determined by the
contributions of such fields to the Noether charge $Q_\xi$ and to the
symplectic potential $\theta$.    Let us first discuss the
anti-symmetric tensor fields.  We restrict to such fields
$A_{a_1\ldots a_p}$ with non-negative mass-squared terms (defined in
Appendix B) and with the standard Maxwell-type kinetic term, noting
that all anti-symmetric tensors in the $S^5$ compactification of
ten-dimensional type IIB supergravity~\cite{AdS5,AdS5krn} to $AdS_5$ and in the
$S^4$ compactification of eleven-dimensional supergravity ~\cite{AdS7}
to $AdS_7$ to satisfy this criterion\footnote{At least, they satisfy
  this criterion to quadratic order in fields.  Under the asymptotic
  behavior specified in appendix B, cubic and higher terms vanish too
  quickly at infinity to contribute.  Such cubic terms arise from
  Chern-Simons terms as well as from couplings to scalar fields.}.
One finds a contribution to $Q_\xi$ proportional to $*F (\xi \cdot A)$
(where $F= dA$) and a contribution to $\theta$ proportional to $*F \wedge \delta A$.

Whether or not these contributions are large enough to affect $\H_\xi$ depends on the asymptotic conditions satisfied by the fields.  As for the case of gravitational fluctuations, we defer a detailed treatment of the linearized equations of anti-symmetric tensor fields to Appendix B.      Under the conditions stated above, Appendix B shows that for a field of given mass and given rank  there is a unique boundary condition such that the evolution operator is self-adjoint in a natural inner product.  For such boundary conditions, the fall-off at inifinity of anti-symmetric tensor fields is too fast to contribute to $\H_\xi$ or to affect the analysis of sections~2 and~3.

Let us therefore examine the case of scalar fields, where the situation is somewhat more subtle.
Consider a scalar field with Lagrange density
\ben
\label{matterL}
L^{\rm matter} = - \frac{1}{2} d^d x \sqrt{-g}[\nabla^a \phi \nabla_a \phi + V(\phi)],
\een
where $V(\phi)$ is a potential\footnote{We assume below that no curvature couplings are present.  The standard presentation of 10- and 11-dimensional supergravity compactified to $AdS_5$ and $AdS_7$ is free of such couplings \cite{AdS5,AdS5krn,AdS7}.
In general, one may choose to work in the Einstein conformal frame and thereby eleiminate any couplings to the Ricci scalar. See, however, \cite{CO} for a treatment of the counter-term subtraction approach in a conformal frame with curvature couplings.}. To quadratic order in $\phi$, it is given by
\ben
V(\phi) = m^2 \phi^2 + \dots.
\een
The prefactor in front of $L^{\rm matter}$ is chosen so that Einstein's equations
for the combined gravity and matter Lagrangian $L = L^{\rm grav} + L^{\rm matter}$ takes the
standard form $R_{ab} - \frac{1}{2} R g_{ab} + \Lambda g_{ab} = 8\pi G \, T_{ab}$, where
the matter stress energy tensor is given by
\ben
\label{phiT}
T_{ab} = \nabla_a \phi \nabla_b \phi - \frac{1}{2} g_{ab}[ \nabla^c \phi \nabla_c \phi + V(\phi)].
\een

Either requiring the energy to be bounded below \cite{bf} or the dynamics to be self-adjoint \cite{iw} restricts $m^2$ to satisfy the so-called Breitenlohner-Friedman bound (BF-bound),
\ben
m^2 \ge -\frac{(d-1)^2}{4\ell^2}.
\een
Scalars arising in 10- and 11-dimensional supergravity compactified to $AdS_5$ and $AdS_7$ have been explicitly shown \cite{AdS5,AdS5krn,AdS7}
to satisfy this bound, though in the $AdS_5$ case scalars do arise that saturate the bound.

For $0 \ge m^2 >  -(d-1)^2/4\ell^2$, solutions to the linearized equations of
motion have the characteristic fall-off at infinity
\begin{equation}
\label{asympt}
\phi \sim \Omega^{\lambda_-} \phi_- + \Omega^{\lambda_+} \phi_+,
\end{equation}
where $\phi_-, \phi_+$ are functions on $\I$, and where
\begin{equation}
\lambda_\pm = \frac{d-1}{2} \pm \frac{1}{2} \sqrt{(d-1)^2 + 4 \ell^2 m^2}.
\end{equation}
When the BF-bound is saturated,
the roots $\lambda_\pm = \lambda$ are degenerate and the second
solution falls off as $\Omega^{\lambda} \log \Omega.$
For masses $m^2 \ge  - (d-1)^2/4\ell^2 +1$, boundedness of the
energy \cite{bf}
or self-adjointness of the evolution \cite{iw} requires the faster
fall-off asymptotic behavior given by $\lambda_+$, but in the range
$- (d-1)^2/4\ell^2 \le  m^2 \le  - (d-1)^2/4\ell^2 +1$ these requirements
impose no restriction.  Thus, for such masses any boundary condition compatible with (\ref{asympt}) may be imposed.

However, even in the range $- (d-1)^2/4\ell^2 \le  m^2 \le -(d-1)^2/4\ell^2 +1$
it turns out that only the most rapidly decreasing asymptotic behavior
(associated with $\lambda_+$ above the BF-bound, and without logs when
the bound is saturated) is compatible with the asymptotic conditions
we imposed on the metric in section 2.  As a result, we restrict our attention
to this rapidly decreasing setting below.
The reader may consult~\cite{cecs1,sb1,sb2,ls,MOTZ,sb3,blmsw,sb4} for recent
work addressing the construction of a conserved energy in settings with
slower fall-off at infinity.

For simplicity, we focus in section (\ref{scalars}) on scalars which
saturate the BF-bound.
We will see that it is marginally compatible with our analysis of the
pure gravitational case.  As a result, scalars with faster fall-off
decrease too rapidly at infinity to affect the expression for
$\H_\xi$, while slower fall-off would require an extended treatment
beyond the scope of the present work.  Rather surprisingly, we will
find that in the marginal BF-bound saturating case the
expression~(\ref{hdef}) for $\H_\xi$ is unmodified.  We will also show in section
\ref{cts} that even in the presence of such fields the counter-term subtraction definition
of energy in $d=5$ continues to differ from $\H_\xi$ by exactly
expression~(\ref{Zdef1}),
just as in the pure gravity case\footnote{For a particular class of black hole solutions \cite{gr} in which scalars of the above form are excited, and for the conformal frame in which $(h_0)_{ab}$ is the metric of the Einstein static universe, it was shown in \cite{ls} that the difference between the counter-term subtraction and Ashtekar et al definitions agrees with this difference evaluated on pure AdS space.  The calculations of \cite{MOTZ,GPP,RT}
also suggest such an agreement.}.

\subsection{Scalar fields saturating the BF-bound and the covariant phase space contruction}
\label{scalars}

Let us now consider in detail scalar fields saturating the BF-bound.
Under our rapid fall-off assumption, these scalars satisfy the following
{\bf asymptotic condition}: In the unphysical spacetime, there is a
function $\phi_0$ which is smooth at $\I$ such that
\ben
\label{BFfast}
\phi = \Omega^{\frac{d-1}{2}} \phi_0.
\een
An immediate consequence of this asymptotic condition
is that the stress energy tensor is of the same order as the electric part of the Weyl
tensor at $\I$. Therefore, since the latter enters in the expression
for $\H_\xi$ in the pure gravity case (see eq.~(\ref{hdef})),
one would now expect there to be an additional contribution
to $\H_\xi$ involving explicitly the scalar field. Surprisingly, a detailed analysis using the same algorithm as above in the pure gravity case shows that such additional contributions are absent, i.e., {\em $\H_\xi$  is given by the same formula in terms of the metric as in the pure gravity case.}\footnote{It would be interesting to investigate whether this continues to be the case under even weaker asymptotic conditions than those explored here.}  Also remarkably, the Hamiltonian
charge $\H_\xi$ is still conserved in the presence of scalars, i.e., it does not depend on the
cross section $C$ on which the integral~(\ref{hdef}) is evaluated.
These results rely on a somewhat subtle cancellation between
various terms, and we therefore now describe the derivation in some detail.

According to the algorithm specified in section~2, we
are instructed to compute the quantities $\xi \cdot \theta(g, \delta g)$, and $\delta Q_\xi$ for an
asymptotic symmetry $\xi^a$, and then seek $\H_\xi$ as a solution to
$\delta \H_\xi = \int_C [\delta Q_\xi - \xi \cdot \theta]$, where $C$
is a cut at infinity, where $Q_\xi$ is the Noether charge of the
theory, and where $\theta$ is the integrand of the surface term
that arises when performing a variation of $L=L^{\rm grav} + L^{\rm
  matter}$ under an integral sign.
In the presence
of matter fields, each of these quantities
generically consists of a gravitational part $Q^{\rm grav}_\xi$ resp.
$\xi \cdot \theta^{\rm grav}$ given by the same formula as in the pure
gravity case (see eqs.~(\ref{td})), as well as contributions
$Q^{\rm matter}_\xi$ resp. $\xi \cdot \theta^{\rm matter}$ from the
matter fields. One has
\ben
\theta_{a_1 \dots a_{d-1}} = \epsilon_{ba_1 \dots a_{d-1}} \left[ \frac{1}{16 \pi G} (
\nabla^c \delta g_c{}^b - \nabla^b \delta g_c{}^c) -
\delta \phi \nabla^b \phi \right]
\een
where the second term is the new contribution from the matter
field. However, for minimally coupled scalars\footnote{
Adding a curvature coupling $L^{\rm curv} = -\alpha d^d x \sqrt{-g} R
\phi^2$ to the matter Lagrangian (\ref{matterL}) would,
in contrast, generate a correction $-2\alpha(\xi^b \phi \nabla^c \phi)
\epsilon_{bca_1 \ldots a_{d-2}}$ to the Noether charge~(\ref{nc}).
Again, this term is finite for scalars satisfying (\ref{BFfast}).
It vanishes for more rapidly decreasing fields and in
general diverges for those falling off more slowly.}  one may show that the
Noether charge
is given by the same expression as in the pure gravity case,
i.e., $Q_\xi$ is given by eq.~(\ref{nc}), with no explicit contributions
from the matter field.

In order to find $\H_\xi$ we must now analyze in more detail the fall off behavior of the metric
and the scalar field $\phi$ near $\I$ using Einstein's equation. In terms of the unphysical metric
$\tilde g_{ab}$, this now takes the form
 \ben
\label{ee1}
 \tilde S_{ab} = -2 \Omega^{-1} \tilde \nabla_a \tn_b + \Omega^{-2}\tilde g_{ab}(\tn^c \tn_c - \ell^{-2}) +
 \tilde L_{ab},
 \een
 where $\tilde S_{ab}$ is given by eq.~(\ref{Sdef}), and
 where $\tilde L_{ab}$ is the matter contribution given by
 \ben
 \tilde L_{ab} \equiv \frac{16 \pi G}{d-2}\left[ T_{ab} - \frac{1}{d-1} g_{ab} T \right].
\label{Labdef}
 \een
 To analyze the consequences of Einstein's equation, we choose, as above, the
 conformal factor so that the unphysical metric takes Gaussian normal form,
 $\tilde g_{ab} = \tilde \nabla_a \Omega \tilde \nabla_b \Omega +
 \tilde h_{ab}$, and perform a power series expansion of
 eq.~(\ref{ee1}), putting $\ell=1$ to simplify the notation.
Einstein's equations give recursion relations for the expansion
coefficients $(\tilde h_{ab})_j$ of the induced metric $\tilde h_{ab}$, the coefficients of the
extrinsic curvature $(\tilde K_{ab})_j$, and the expansion coefficients
of the scalar field, $\phi = \Omega^{(d-1)/2}( \phi_0 + \Omega \phi_1 + \dots)$, which can
be solved in a manner similar to the pure gravity case in Sec.~3. By an argument
completely analogous to that given in the pure gravity case, one finds that the coefficients
$(\tilde h_{ab})_j$ for $j<d-1$ and $(\tilde K_a{}^b)_l$ for $l<d-2$
are the same as for the exact AdS metric, while the coefficients
$(\tilde h_{ab})_{d-1}$ and $(\tilde K_a{}^b)_{d-2}$ are
related to the leading order electric Weyl tensor and the
leading order coefficient of the scalar
field\footnote{It is important to note here that the recursion
relation eq.~(\ref{recursionp}) at order $j = d-2$ provides a non-trivial consistency check for the asymptotic condition on the scalar field, because it says that the right side (supplemented with
a term proportional to the trace-free part of $\tilde h_c{}^a \tilde h_b{}^d \tilde L_a{}^b$) must vanish. This turns out  to be the case, since  $\tilde h_c{}^a \tilde h_b{}^d \tilde L_a{}^b$
at this order is a pure trace (see eq.~(\ref{phiT})), and because the gravitational contributions
cancel as well.}.   On the other hand, the higher coefficients $\phi_1, \phi_2, \dots$
and $(\tilde h_{ab})_j$ for $j > d-1$ and $(\tilde K_a{}^b)_l$ for $l > d-2$
are uniquely determined by the recursion relations and the leading
order electric Weyl tensor, as well as the leading scalar coefficient $\phi_0$.

In order to derive the relation between $(\tilde h_{ab})_{d-1},
\phi_0$ and the leading order electric Weyl tensor, one proceeds in
the same manner as in the pure gravity case. The arguments leading up
to eq.~(\ref{A-Key-equation-2}) now result in
\ben
\label{CKL}
 \tilde C_{abcd} \tilde n^b \tilde n^d =
{\pounds_{\tilde n} \tilde K_{ac}} - \Omega^{-1} \tilde K_{ac}
 + \tilde K_{ab} \tilde K^b{}_c - \frac{1}{2} \tilde L_{bd} \tilde h_a{}^b \tilde h_c{}^d
 - \frac{1}{2} \tilde h_{ac} \tilde L_{bd} \tn^b \tn^d
\een
which differs only by an additional matter term involving the
matter stress tensor $\tilde L_{ab}$. From the asymptotic condition on
$\phi$, this has an expansion of the form
 \ben
 \tilde L_{ab} = \Omega^{d-3}(\tilde L_{ab})_{d-3} + \Omega^{d-2}(\tilde L_{ab})_{d-2} + O(\Omega^{d-1}),
 \een
 where
 \bena
 (\tilde L_{ab})_{d-3}  &=& 4 \pi(d-1) G \left[ \phi_0^2 \tn_a \tn_b
 - \frac{1}{d-2} \phi_0^2 \tilde h_{ab} \right] \label{Lexp1}\\
 (\tilde L_{ab})_{d-2} &=& \frac{8 \pi G}{d-2} \left[
 2(d-1) \phi_0 \tn_{(a} \tilde \nabla_{b)} \phi_0 - d \cdot \phi_0 \phi_1 \tilde g_{ab}
 \right].
\label{Lexp2}
 \eena
We substitute these expressions into eq.~(\ref{CKL}), and we expand the resulting equation in
powers of $\Omega$, as we did above in the pure gravity case. At each order in $\Omega$, this then gives a relation between the expansion
coefficients of the electric Weyl tensor, the coefficients $(\tilde h_{ab})_j$
of the induced metric, the coefficients of the
extrinsic curvature $(\tilde K_{ab})_j$, and the expansion coefficients $\phi_j$
of the scalar field. At order $j = d-1$,
and in dimensions $d=4, d \ge 6$, the relation can be brought into the form
\ben
(\tilde h_{ab})_{d-1} = -\frac{2}{d-1} (\tilde E_{ab})_0 -
\frac{4 \pi G}{d-2} (\tilde h_{ab})_0 \phi_0^2,  \quad {\rm for} \,\, d=4, d\ge 6,
\een
where $\tilde E_{ab}$ is the leading order electric Weyl tensor given
by eq.~(\ref{Eabdef}). In $d=5$,
the corresponding relation turns out to be
\ben
(\tilde h_{ab})_{4} = -\frac{1}{2} (\tilde E_{ab})_0 - \left( \frac{1}{16} +  \frac{4 \pi G \, \phi_0^2 }{3} \right) (\tilde h_{ab})_0  \quad {\rm for} \,\, d=5.
\een
One immediately finds from this that unphysical line element must take the form\footnote{Here for simplicity we have as usual assumed that $(\tilde h_{ab})_0$ is the standard metric \ref{pureads} on the Einstein Static Universe so that the conformal factor satisfies (\ref{noshift}).}
\bena
\label{ule6}
d \tilde s^2 &=& d\Omega^2 -
\left[ \left(1 + \frac{1}{4} \Omega^2 \right)^2 - \frac{4\pi G \phi_0^2}{d-2} \Omega^{d-1} \right]\, dt^2
+ \left[ \left( 1 - \frac{1}{4} \Omega^2 \right)^2 -\frac{4\pi G \phi_0^2}{d-2} \Omega^{d-1} \right] \, d\sigma^2 \nonumber\\
&& - \Omega^{d-1} \frac{2}{d-1} \, \tilde E_{ij} \, dx^i dx^j
+ O(\Omega^{d})\quad {\rm for}\,\, d\ge 5.
\eena
In $d=4$, the corresponding result is
\bena
\label{ule45}
d \tilde s^2 &=& d\Omega^2 - 
\left(1 + \frac{1}{2} \Omega^2 - 2\pi G\, \phi_0^2 \Omega^{3} \right) \, dt^2
+ \left( 1 - \frac{1}{2} \Omega^2 - 2 \pi G \, \phi_0^2 \Omega^{3} \right) \, d\sigma^2 \nonumber\\
&& - \frac{2}{3} \Omega^3 \, \tilde E_{ij} \,
dx^i dx^j + O(\Omega^{4})\,  \quad {\rm for}\,\, d=4.
\eena

The rest of the analysis is now similar to the pure gravity case. From
the asymptotic form of the metric given above, one finds that the variation of the
Noether charge $\delta Q_\xi$ has exactly the same form as in the pure gravity case~(\ref{delqu}),
even though the metric feels the backreaction of the scalar field, as seen in eqs.~(\ref{ule6}, \ref{ule45}).
The contributions from the scalar field in these expressions happen to cancel out in $\delta Q_\xi$.
One furthermore calculates that
\ben
\delta \phi \nabla^b \phi = \Omega^{d}\left( \frac{d-1}{2}\tilde n^b \phi_0 \delta \phi_0 + O(\Omega) \right),
\een
and that
\ben
\frac{1}{16\pi G}(\nabla^c \delta g_c{}^b - \nabla^b \delta g^c{}_c) = \Omega^{d}\left(
\frac{d-1}{2}\tilde n^b \phi_0 \delta \phi_0 + O(\Omega) \right)
\een
It follows from these expressions
that the terms involving the contributions involving $\delta \phi_0$ from the matter field
precisely cancel out also in
the expression $\xi \cdot \theta$.
Consequently, $\delta \H_\xi$ receives
no explicit contributions from the matter field and is
therefore given by exactly the same expression as in the pure gravity
case. Thus
\ben
\label{Hdefscalar}
\H_\xi = \frac{-1}{8 \pi G} \,\,  \int_C
\tilde E_{ab} \tilde u^b \xi^a \, d \tilde S.
\een

Let us finally show that the hamiltonian charges $\H_\xi$ continue to
be conserved in the presence of
a scalar field saturating the BF-bound. As in the pure gravity case, we may use Gauss' law to
express the difference between the generators defined on two different cuts $C_1, C_2$ as
the integral over the enclosed portion $\I_{12}$ of scri of the
divergence of the integrand $\tilde E_{ab} \xi^b$, see eq.~(\ref{balance}).
In the pure gravity case, the divergence was shown to be zero.
We now show that this is still the case in the presence of the scalar
field $\phi$. This again follows from Einstein's equation, which in the presence of matter fields
leads to the relation
\ben
\tilde \nabla^d \left( \frac{\Omega^{3-d}}{d-3} \tilde C_{abcd} \right) = -
\Omega^{2-d} [\tilde L_{d[a} \tilde g_{b]c} \tilde n^d + \tilde\nabla_{[a}(\Omega \tilde L_{b]c})]
\een
for any conformal factor $\Omega$.
Now contract both sides of the equation into $\tn^b \tn^c \tilde h^a{}_e$, and take the limit
as $\Omega \to 0$. On the right side, we get one potentially diverging term coming
from $(\tilde L_{ab})_{d-3}$ and one converging term coming from $(\tilde L_{ab})_{d-2}$
and from $\tilde \nabla_{[a} (\tilde L_{b]c})_{d-3}$.
However, the diverging term is seen to vanish using eq.~(\ref{Lexp1}) and taking into
account the contractions, while the contributions to the
converging term all happen to cancel each other using
eqs.~(\ref{Lexp1}, \ref{Lexp2}).
Thus, the limit of the
contraction of the right side as $\Omega \to 0$ vanishes.
The left side gives $\tilde D^a \tilde E_{ea}$.
Thus, $\tilde D^a \tilde E_{ea} = 0$ on $\I$, showing that the charges
$\H_\xi$ are conserved in the presence of the matter field $\phi$.

\subsection{Scalars saturating the BF-bound and the counter-term
subtraction notion of energy}
\label{cts}

From the results above, it is clear that the covariant phase space
definition of energy continues to agree with the Ashtekar et al definition
of energy in the presence of scalars satisfying the asymptotic
condition~(\ref{asympt}). Furthermore, since our asymptotic conditions on
the metric (from section \ref{approach}) continue to hold, the arguments
of section \ref{htsubsec} once again show that the covariant phase space
charge also agrees with the Henneaux-Teitelboim definition.

Let us now compare this definition with the counter-term subtraction method.
For scalars saturating the BF-bound (and integer steps above this bound), the counter-term subtraction scheme was analyzed in \cite{KS6,KS7,KS8,KS9}, but we repeat the analysis of the BF-saturating case here for completeness and for consistency with our current notation.

For scalars satisfying (\ref{asympt}), the scalar field counter-term
required to make the scalar action finite and to yield a well-defined
variational principle is given by e.g.~\cite{KW,Minces,SevShom,Lorentz},
\ben
- \frac{d-1}{4} d^{d-1}x \sqrt{-h}\phi^2.
\een
Thus, the combined effective boundary stress-energy tensor is the sum
of the gravitational boundary stress tensor and the stress tensor
associated with the scalar boundary Lagrangian. In $d=5$, this is
\ben
\tau_{ab} = \frac{1}{8 \pi G} \left[ \frac{1}{2}\left( \R_{ab} - \frac{1}{2} \R h_{ab} \right)
- 3 h_{ab} - K_{ab} + K h_{ab} \right] - h_{ab} \phi^2
\een
and the counterterm subtraction charge $\Q_\xi$ in the full theory is again given by
eq.~(\ref{qcounterdef}). Let us compare $\Q_\xi$ with the charge $\H_\xi$,
obtained in the covariant phase space approach. As shown in the
previous subsection, the charge $\H_\xi$ is given by exactly the
same expression as in the pure gravity case, see eq.~(\ref{hdef}),
and does not contain any explicit contributions from the scalar
field. On the other hand $\tau_{ab}$, and hence $\Q_\xi$, does contain
an explicit contribution from the scalar field. Thus, one would
naively expect that the difference between $\Q_\xi$ and $\H_\xi$ would
depend on the boundary value of the scalar field under consideration.
However, as we will now show, this is not the case, and the difference
$\Q_\xi - \H_\xi$ is given by exactly the same expression as in the
pure gravity case, see eq.~(\ref{Zdef}).  This difference is therefore just a
constant offset which does not depend on the particular field
configuration under consideration, but only on $\xi^a$ and
``kinematical'' data fixed by the asymptotic conditions.

Let $\I_\Omega$ be the timelike surfaces of constant $\Omega$,
let $h_{ab}, {\mathcal R}_{abcd}$
and $K_{ab} = -h_a{}^c h_b{}^d \nabla_c \hat \eta_d$
be the induced metric, the intrinsic
Riemann tensor, and the extrinsic curvature, and let $\hat \eta^a$ be the
unit normal to $\I_\Omega$. Then, from the Gauss-Codacci equations
and Einstein's equation, we have
\ben
{\mathcal R}_{ac} = L_{bd} h^b{}_a h^d{}_c + \frac{1}{2} L h_{ac} -
\frac{1}{2} h_{ac} L_{bd} \hat \eta^b \hat \eta^d - C_{abcd} \hat
\eta^b \hat \eta^d + K K_{ac} - K_{ab} K^b{}_c - 3h_{ab},
\een
where $L_{ab} = \tilde L_{ab}$ is the matter contribution given by
eq.~(\ref{Labdef}). Now express the quantities on the right side of this equation
in terms of the corresponding unphysical quantites which are given by
$h_{ab} = \Omega^{-2} \tilde h_{ab}$, $K_{ab} =
\Omega^{-1} \tilde K_{ab} + \Omega^{-2} \tilde h_{ab}$,
$\tilde n^a = \tilde \nabla^a \Omega$, and substitute the resulting
expression for ${\mathcal R}_{ac}$ into the
definition of $\tau_{ab}$. Then one obtains
\bena
\tau_{ab} - \frac{-1}{16 \pi G} C_{acbd}
\hat \eta^c \hat \eta^d &=&
\frac{-1}{16 \pi G} \Bigg[ -\tilde K \tilde K_{ab} + \tilde K_a{}^c
\tilde K_{cb} - \frac{1}{2}(- \tilde K^2 + \tilde K_{cd} \tilde
K^{cd}) \tilde h_{ab} \nonumber \\
&+& \tilde L_{cd} \tilde h^c{}_a \tilde h^d{}_b - \tilde L \tilde h_{ab}
+ \tilde h_{ab} \tilde L_{cd} \tilde n^c \tilde n^d \Bigg] -
\Omega^2 \phi^2_0 \tilde h_{ab},
\eena
where we are assuming for simplicity that $\Omega$ has been chosen
so that $\tilde n^a \tilde n_a = 1$. Now, using the expansion~(\ref{Lexp1})
of $\tilde L_{ab}$, we see that the matter terms in the last line
precisely cancel (up to order $\Omega^{3}$). Contracting the above
equation into $\xi^a \tilde u^b$ (with $\tilde u^a$ the unit
timelike normal to a cut $C$ of $\I$), dividing by $\Omega^2$, and
integrating over $C$, we therefore find that $\H_\xi - \Q_\xi$ is given
by exactly the same expression as in the pure gravity case, eq.~(\ref{qhdiff}).
As in the pure gravity case, this can be brought into the final
form~(\ref{Zdef}) using eq.~(\ref{kabrab}) (which does not receive any additional
contributions from the scalar field at the relevant order).

\subsection{... and the spinor charge}
\label{spinorchargesec}

Let us finally compare the covariant phase space charges with the
spinor charge in the presence of a scalar field $\phi$ with
a potential $V$, saturating the BF-bound. In the pure gravity case,
the relevance of the spinor charge is that it has a positivity
property. As shown in \cite{t,boucher}, this positivity property continues
to hold in the presence of a single scalar field $\phi$ if and only
if the potential $V$ arises from a superpotential $W$. More precisely, let
\ben
P(\phi) = -\frac{1}{8\pi G}(d-1)(d-2) + V(\phi)
\een
be the combined scalar field potential and
contribution from the cosmological constant (with $\ell = 1$). Then
there must be a $W$ such that
\ben
\label{Pdef}
8\pi G \, P = -4(d-2)(d-1)W^2 + 4(d-2)^2 (\partial W/\partial \phi)^2.
\een
The spinor charge is then defined by the same formula as
in the pure gravity case [see eq.~(\ref{spinorcharge})],
but with the $\widehat \nabla_a$-operator
now defined by
\ben
\label{hatdef}
\widehat \nabla_a \psi = \nabla_a \psi - W(\phi)\gamma_a \psi.
\een
The condition that $P$ be given in terms of $W$ by eq.~(\ref{Pdef}),
and that $V = -\frac{1}{4}(d-1)^2 \phi^2 + \dots$ for a
field saturating the BF-bound leads to
\ben
\label{Wasympt}
W = \frac{1}{2} + \frac{\pi G(d-1)}{(d-2)} \phi^2 + \dots \, .
\een
Let $\psi$ be a Killing spinor in exact AdS space
(with $\phi = 0$), and let $\xi^a = \overline \psi \gamma^a \psi$.
For an asymptotically AdS metric $g_{ab}$
and a scalar field $\phi$ with the asymptotic behavior $\phi =
\Omega^{(d-1)/2} \phi_0$ satisfying Einstein's equation,
the asymptotic expansion~(\ref{ule6}) of the
metric holds. From this, and from eq.~(\ref{spinorid}), one obtains
\ben
\nabla_a \psi - \frac{1}{2} \gamma_a \psi
= \Omega^{d-5/2} \tilde \gamma^b \tilde
\gamma^c \left[
\frac{1}{2} \tilde E_{a[b} \tilde n_{c]} + \frac{\pi G(d-1)}{d-2}
\phi_0^2
\tilde h_{a[b} \tilde n_{c]}
\right] \tilde \psi + O(\Omega^{d-3/2}).
\een
Now let $\hat u^a$ be the unit future timelike normal to a cross section
$C$ in $\I$. Substituting the above expression
into the definition of the Nester 2-form
$B_{ab}$ and using eqs.~(\ref{Wasympt}, \ref{hatdef}) gives
\ben
\Omega^{2-d} B_{ab} \hat \eta^a \hat u^b = \frac{-1}{8\pi G} \tilde
E_{ab} \tilde u^a \xi^b + O(\Omega),
\een
where $\tilde u^a = \Omega \hat u^a$. Thus, the explicit contribution
from the scalar field entering the spinor charge via
eq.~(\ref{hatdef}) is precisely canceled by the indirect
contribution of the scalar field to the metric via Einstein's equation.
Integrating over $C$ gives that the spinor charge $\Q_\xi$ again agrees
with the Hamiltonian charge $\H_\xi$ given by eq.~(\ref{Hdefscalar})
in the presence of scalar fields. From the positivity of
the spinor charge we therefore find that $\H_\xi$ also satisfies
the positive energy theorem, provided the the scalar field potential
arises from a superpotential.

\section{Perturbation analysis of the asymptotic behavior of
the gravitational field}

In the previous sections, we have worked out the consequences of our asymptotic
conditions on the metric. The AdS-group is the asymptotic symmetry
group of these conditions, and we showed that gravitational charges associated with asymptotic
symmetries can be defined in a clear-cut and natural way, and that the resulting
expression agrees with expressions proposed previously in the literature.
While these results lend support\footnote{Note that a consistency check of the key
assumption that the unphysical metric be smooth (i.e., $C^\infty$) was implicit
in writing it as a formal power series in $\Omega$ in section~3.
The key point is that the recursion relations fixing the coefficients of this power
series can be solved consistently without the need to, say, include logarithmic
terms in the expansion.} to our choice of asymptotic conditions, they are
not, of course, a proof that a ``generic'' metric will satisfy these
conditions, i.e., that there is a sufficiently wide class of solutions
satisfying these conditions, nor are they a proof that our asymptotic
conditions are the only possible ones.

Unfortunately, due to the nonlinearities of Einstein's equations,
it is difficult to analyze this issue in any straightforward manner. Nevertheless,
it is possible to address it in the context of perturbation theory, and we shall do so
in this section. At the level of perturbation theory, the corresponding question is
as follows:
Let $\gamma_{ab}$ be a solution to the linearized Einstein equation about exact AdS space.
Can one extend the perturbed unphysical metric to $\I$ so as to be
smooth (or suitably many times differentiable) there modulo
a gauge transformation $\pounds_\eta g_{ab}$, and a change $\delta \Omega$ of the conformal
factor $\Omega$ of exact AdS space?
Thus, given $\gamma_{ab}$, we ask whether there is a vector field $\eta^a$ on pure AdS
space, and a function $\delta \Omega$ (with $\delta \Omega = 0$ on $\I$), such that
\ben
\tilde \gamma_{ab} = \Omega^2 \psi^* (\gamma_{ab} + \nabla_a \eta_b + \nabla_b \eta_a)
+ \Omega \delta \Omega \psi^* g_{ab}
\een
is smooth and vanishing at $\I$, where $\psi$ is some diffieomorphism.
(Here and in the remainder of this section, $g_{ab}$ is the metric of exact AdS-space,
and $\nabla_a$ is the derivative operator in exact AdS space.)
Actually, rather than analyze the perturbed metric, it is much more convenient to analyze
instead the perturbed Weyl tensor, $\delta C_{abcd}$. This has the triple advantage
that the perturbed Weyl tensor is both gauge invariant and conformally
invariant, and that it is the key quantity of interest in our formula for
the (perturbed) charges of the spacetime.  If the Weyl tensor with one upper index
$\delta C_{abc}{}^d$ is sufficiently
smooth on the unphysical spacetime, then by our previous analysis of the Einstein equations the unphysical metric will be sufficiently smooth.
Consequently, by working with $\delta C_{abcd}$, we are immediately
able to determine whether a generic metric perturbation gives rise
to a space time for which the charges $\H_\xi$ can still be defined
(to first order).

Using the linearized Einstein's equations and the Bianchi identities,
it can be shown that the perturbed Weyl tensor off of pure AdS-space
satisfies the following wave equation:
\ben
\label{waveeqweyl}
\left( \nabla^e \nabla_e + \frac{2(d-1)}{\ell^2}\right) \delta C_{abcd} = 0 \, .
\een
To understand the consequences of this wave equation, we consider the field
\ben
\Psi \equiv r^{(d-2)/2} Y^{-2} \delta C_{abcd} (\nabla^a Y) t^b (\nabla^c Y) t^d
\een
where $t,r$ are the standard global time and radial coordinates of AdS,
see eq.~(\ref{pureads}), $t^a = (\partial/\partial t)^a$, and where
$Y = \sqrt{1 + r^2/\ell^2} \cos(t/\ell)$. Thus, at $\I$, $\Psi$ is
essentially the $trtr$-component of the perturbed Weyl tensor
multiplied by a power of $r$.  The other components of the Weyl tensor are addressed in Appendix~A with similar results.

Now make a Fourier decomposition of $\Psi$ into modes with time
dependence $\exp(-i\omega t/\ell)$, and with angular dependence given
by a spherical harmonic on $S^{d-2}$ with angular momentum quantum
number $l$. As we show in Appendix~A, from eq.~(\ref{waveeqweyl}),
each such mode obeys the ordinary differential equation
\ben
  \left(
       - \frac{\partial^2}{\partial x^2}
       + \frac{\nu^2 - 1/4}{\sin^2 x}
       + \frac{\sigma^2-1/4}{\cos^2 x}
  \right) \Psi =  \omega^2 \Psi \,,
\label{eq:eigenvalue}
\een
where $\sigma = l + (d-3)/2 \,, \quad l=2,\,3,\, \dots$, where
$\nu = |d-5|/2$, and where a radial coordinate $x$ has been defined by
\ben
\frac{\ell}{r} = \tan x,
\een
so that $x=0$ represents points at infinity, while $x = \pi/2$ represents
the origin of polar coordinates $r=0$. We note that the equation satisfied
by $\Psi$ has exactly the same form as the ``master equation'' for
the metric perturbations off of pure AdS space found in~\cite{iw},
where a detailed analysis of that equation can be found.
The two linearly independent solutions to (\ref{eq:eigenvalue}) are
given in terms of hypergeometric functions $F = {}_2F_1 $.
Different combinations of those solutions correspond to different behaviors of $\Psi$,
and hence of the corresponding Weyl tensor component, at infinity $x=0$
and at the interior point $x= \pi/2$.

With the AdS-CFT correspondence in mind, we restrict attention to solutions
that are regular inside the AdS-bulk in the sense that no boundary terms
from the interior region of AdS appear in action integral.
This regularity requires each mode function $\Psi$ to be vanishing
and normalizable at the origin\footnote{ 
For this regularity, each mode function $\Psi$ itself need not be smooth
at $x=\pi/2$. Indeed, for $d$-odd case, each normalizable mode
function $\Psi$ fails to be smooth at $x=\pi/2$.
If one considers a timeslice $\Sigma$ with a single point
removed---as considered in some cases of interest,
e.g., a conical singularity as a particle in AdS---then the singlar
solution may be allowed and come to play some role in AdS-CFT correspondence.
Also when some matter fields and/or black holes exist inside the bulk
so that non-linear effects become essential, the situation may change.
} 
$x=\pi/2$. Using the fact that $\sigma
\ge 1$, one finds that the solutions with this property are given by
\ben
 \Psi \propto
 (\sin x)^{\nu + 1/2} (\cos x)^{\sigma +1/2}
 F\left( \zeta^\omega_{\nu,\sigma}, \zeta^{-\omega}_{\nu,\sigma},
  1+\sigma; \cos^2 x \right) \,,
\label{sol:smooth}
\een
where
\ben
 \zeta^\omega_{\nu,\sigma} \equiv \frac{\nu + \sigma +1 +\omega}{2} \,.
\een
In order to see the asymptotic behavior of this solution near infinity $x=0$,
it is convenient to transform the argument $\cos x$ into $\sin x$
using well-known transformation identities for hypergeometric functions.
For the even-$d$-case, one obtains
\bena
 && \Psi \propto (\cos x)^{\sigma +1/2}
        \Bigg\{
         \frac{\Gamma(1+\sigma) \Gamma(\nu)
                }{
                  \Gamma(\zeta_{\nu,\sigma}^{\omega})
                  \Gamma(\zeta_{\nu,\sigma}^{-\omega})
                }
           (\sin x)^{-\nu + 1/2}
           F\left(
                  \zeta_{-\nu,\sigma}^{\omega},
                  \zeta_{-\nu,\sigma}^{-\omega},
                  1-\nu; \sin^2 x
           \right)
 \non \\
 && \qquad \quad \,\,
      +    \frac{\Gamma(1+\sigma) \Gamma(- \nu)
                }{
                  \Gamma(\zeta_{-\nu,\sigma}^{\omega})
                  \Gamma(\zeta_{-\nu,\sigma}^{-\omega})
                }
           (\sin x)^{\nu + 1/2}
           F\left(
                  \zeta_{\nu,\sigma}^{\omega},
                  \zeta_{\nu,\sigma}^{-\omega},
                  1+\nu; \sin^2 x
            \right)
      \Bigg\} \,.
\label{transf:hypergeometric-nu}
\eena
For the odd-$d$-case, if $\omega$ is {\em not} a real value such that
\ben
 \omega = \mp \left( 2m + 1 + \nu + \sigma \right) \,,
 \quad m = 0,\,1,\,2,\, \cdots \,,
\label{condi:quantization}
\een
we have
\bena
&&\Psi \propto (\cos x)^{\sigma +1/2}
    \Bigg[\;
      \frac{\Gamma(1+\sigma) \Gamma(\nu)
          }{\Gamma(\zeta_{\nu,\sigma}^{\omega})
            \Gamma(\zeta_{\nu,\sigma}^{-\omega})
          }
          \sum_{k=0}^{\nu -1}
          \frac{(\zeta_{-\nu,\sigma}^{\omega})_k
                (\zeta_{-\nu,\sigma}^{-\omega})_k
               }{k! (1-\nu)_k } \cdot
          (\sin x)^{-\nu + 1/2 + 2k}
\non \\
  && \quad \,\,
   - \frac{(-1)^{\nu}\Gamma(1+\sigma)
          }{\Gamma(\zeta_{-\nu,\sigma}^{\omega})
            \Gamma(\zeta_{-\nu,\sigma}^{-\omega})
          }
          \sum_{k=0}^{\infty}
          \frac{(\zeta_{\nu,\sigma}^{\omega})_k
                (\zeta_{\nu,\sigma}^{-\omega})_k
               }{k!(\nu+k)!} \cdot
          (\sin x)^{\nu + 1/2+ 2k} \cdot
          \Big\{ \log (\sin^2 x)
\non \\
 &&  \qquad
                 -\psi(k+1)-\psi(k+\nu+1)
                 + \psi(\zeta_{\nu,\sigma}^{\omega} + k)
                 + \psi(\zeta_{\nu, \sigma}^{-\omega} + k)
          \Big\} \;
     \Bigg] \,.
\label{transf:hypergeometric-integer}
\eena
where $(\zeta)_k \equiv \Gamma(\zeta+k)/\Gamma(\zeta)$ and
$\psi$ is the di-gamma function.
Note that for the $d=5$ case, $\nu =0$ and the first term on
the right-hand side of eq.~(\ref{transf:hypergeometric-integer})
should be ignored.

From these expressions, we immediately see that
when $0 \leq \nu <1$, i.e. in the cases $d=4,\,5,\,6$,
$\Psi$ is normalizable at $x=0$ for any $\omega \in {\Bbb C}$.
In this case, expanding the above expressions
(eq.~(\ref{transf:hypergeometric-nu}) with $\nu=1/2$ and
eq.~(\ref{transf:hypergeometric-integer}) with $\nu=0$) in $\sin x$,
we find that for $d=4, \,6$,
\ben
\Psi = A  + B \sin x + C (\sin x)^2 + \dots
\label{asympt:46}
\een
and for $d=5$,
\ben
\Psi = (\sin x)^{1/2} [ A  \log(\sin x) + B ][1 + C (\sin x)^2 + \cdots ],
\label{asympt:5}
\een
where `$\cdots$' represents a convergent power series in $\sin x$ and
the coefficients $A,\,B,\,\dots$ depend on the parameters $\nu$, $\sigma$,
and $\omega$. In particular, the ratio $B/A$, which takes an arbitrary
real number, specifies all possible boundary coinditions that can
be imposed at $x=0$ \cite{iw}.


In the case when $d \geq 7$, the solution $\Psi$ is not
normalizable at $x=0$ when $\omega$ fails to satisfy the
``quantization condition'' eq.~(\ref{condi:quantization}). On the
other hand, when the quantization condition is satisfied, then
one can determine\footnote{ 
When eq.~(\ref{condi:quantization}) holds, either $\zeta^\omega_{\nu,\sigma}$
or $\zeta^{-\omega}_{\nu,\sigma}$ becomes zero or a negative integer, $-m$,
and hence the first term on the right-hand side of
eq.~(\ref{transf:hypergeometric-nu}) vanishes. In particular,
for $\nu \in {\Bbb N}$, i.e., $d=7,\,9,\,11,\,\dots$,
$\zeta^\omega_{-\nu,\sigma}$ (or $\zeta^{-\omega}_{\nu,\sigma}$) also
is a negative integer, $-m-\nu$, hence the expression,
eq.~(\ref{transf:hypergeometric-integer}), itself becomes trivial
and does not make sense.
} 
the asymptotic behavior at $x=0$ by directly
examining the solution, eq.~(\ref{sol:smooth}),
instead of using the expressions~(\ref{transf:hypergeometric-nu})
and~(\ref{transf:hypergeometric-integer}).
In fact, the qunatization condition~(\ref{condi:quantization}) implies
$\zeta^\omega_{\nu,\sigma}=-m$ (or otherwise
$\zeta^{-\omega}_{\nu,\sigma}=-m$), hence $F(\zeta^\omega_{\nu,\sigma},
\zeta^{-\omega}_{\nu,\sigma},1+\sigma;\cos^2 x)$ becomes a polynomial of
$\cos x$ of order at most ${2m}$. Therefore the asymptotic behavior of
$\Psi$ is simply given by the factor $(\sin x)^{\nu+1/2}$, i.e.,
\ben
 \Psi = (\sin x)^{(d-4)/2} \left[1 + C \sin x + \cdots \right] \quad (d \ge 7) \, .
\label{asympt:78910}
\een
To see explicitly how the behavior of $\Psi$ which we have just
found relates to the behavior of the corresponding Weyl component
at $\I$, we choose the conformal factor as $\Omega = \sin x$.
Then $\Omega^2$ times the exact AdS metric is smooth at $\I$, and
\ben
\delta \tilde E_{ab} t^a t^b \sim \Omega^{-(d-4)/2} \Psi, \quad
{\rm as} \,\, \Omega \to 0 \,.
\een
The asymptotic formulae eqs.~(\ref{asympt:46}), (\ref{asympt:5}), and
(\ref{asympt:78910}), translate into the following asymptotic behavior
of the perturbed electric Weyl tensor:
\ben
\begin{array}{cc}
d & \delta \tilde E_{ab} t^a t^b\\
4 & A + B\Omega + \cdots\\
5 & (A \log \Omega + B)(1 + C\Omega + \cdots)\\
6 & A \Omega^{-1} + B + C \Omega + \cdots \\
\geq 7 & 1 + C \Omega^2 + \cdots
\end{array}
\een
Thus, we arrive at the following conclusion: If we demand
normalizability of the perturbed Weyl tensor in the interior,
then in  $d=4$ and in $d\geq 7$, the perturbed electric Weyl tensor
component $\delta \tilde E_{ab} t^a t^b $ is smooth at $\I$.
This justifies our boundary condition for asymptotic AdS-spacetime
within linear perturbation analysis in these dimensions.
In $d=5$, it is smooth if $A=0$, but it is logarithmically divergent
if $A \neq 0$, while in $d=6$, it is smooth if $A=0$ and diverges
as $\Omega^{-1}$ if $A \neq 0$.


As mentioned above, in $d=4, 5, 6$, the ratio $B/A$
(with $\pm \infty$ allowed) corresponds to a specific choice of boundary
conditions for the wave equation~(\ref{waveeqweyl}).
In $d=4$ there is hence a one-parameter family of boundary conditions
giving rise to a smooth perturbed electric Weyl tensor at $\I$. On the other
hand, in $d=5, \, 6$, there is only one choice ($A=0$, essentially
``Dirchlet conditions'') corresponding to
a smooth electric Weyl tensor. The freedom of choosing boundary
conditions in the cases $d=4, 5, 6$ suggests that there are other
notions of asymptotically AdS spacetimes in those cases, with
consequently different expression for conserved quantities associated
with the asymptotic symmetries.
On the other hand, in spacetime dimension $d \ge 7$,
the boundary conditions are unique, at least at the level of
linear perturbation theory.


Actually, as shown in Appendix~A, the other electric-type
components of the perturbed Weyl tensor, such as
$\delta \tilde E_{ab}t^a (\partial/\partial \theta^i)^b$,
$\delta \tilde E_{ab}(\partial/\partial \theta^i)^a
(\partial/\partial \theta^j)^b$, also have the same asymptotic behavior
as $\delta \tilde E_{ab}t^at^b$.  Furthermore, all other components are determined from the electric components by the symmetries of the Weyl tensor,
the Bianchi identity, and Einstein's equation.
As a result, the above discussion of boundary conditions and
the linearized regularity problem of the conformal AdS-infinity $\I$
applies to all types of the Weyl curvature perturbations.

\section{Summary}
\label{disc}

In this work we have compared various constructions of conserved charges (e.g., energy) in asymptotically anti de-Sitter spaces, and also introduced a new construction following the covariant phase space method of Wald et al. \cite{wz,wi}.    Our main results are as follows:

\begin{itemize}
\item  In $d \ge 4$ spacetime dimensions the Ashtekar et al definition \cite{asht1,asht2} based on the electric part of the Weyl tensor agrees with the Hamiltonian construction due to Henneaux and Teitelboim \cite{ht} and with the covariant phase space definition under suitable asymptotic  conditions.  These conditions are stated in  detail in section \ref{approach}.

\item  This agreement occurs because our asymptotic conditions guarantee that the expansion of any asymptotically AdS metric near infinity can be written in the simple form (\ref{unphysmetric:678911},\ref{unphysmetric:45}) in terms of the electric part of the Weyl tensor.

\item In $d=5$ the above definitions of conserved charges differ from the charges defined by the counter-term subtraction method \cite{skenderis,kraus}.  However, this difference is a function only of the auxiliary conformal structure at ${\I}$ required by the counter-term subtraction method.  This agrees with previous results (e.g., \cite{skenderis,kraus,KS2,KS3,KS4,KS5,KS6,KS7,KS8,KS9}) which evaluated this difference on particular solutions (e.g., Schwarzschild-AdS).  Note that, as a result, the counter-term energy is also consistent with the
covariant phase space methods of \cite{wz}, which controls only variations
of the Hamiltonian on the space of solutions.
Equation (\ref{Zdef}) displays the explicit formula giving this difference for a general conformal structure.  Because this difference is constant over phase space, it has trivial Poisson bracket; that is, the charges generate the same Hamiltonian vector fields.  The choice to study the $d=5$ was made for simplicity.  See \cite{peierls} for an argument based on the Peierls bracket to the effect that a similar conclusion must hold in all dimensions.

\item A linearized analysis near infinity and consideration of the resulting back-reaction indicates that our asymptotic conditions are consistent with the dynamics of the metric, of anti-symmetric tensor fields and vector fields with $m^2 \ge 0$, and of scalar fields with $m^2$ either above or saturating the Breitenlohner-Friedman bound.
\end{itemize}

We should also emphasize those points not investigated here.
For example, our analysis was confined to cases with spacetime dimension
$d \ge 4$. The $d=3$ case will require special treatment due to the frequent
appearance of factors of $d-3$ in section~\ref{s3}. Furthermore,
in $d=3$ spacetime dimensions the Weyl tensor vanishes identically so that
the Ashtekar et al definition becomes trivial, though one would still
expect the other constructions to agree in the manner described above.

One could also weaken the definition of asymtotically anti-de
Sitter in several ways.  For example, in the AdS/CFT context it is
common to consider spacetime satisfying asymptotic conditions
similar to those stated in section~\ref{approach}, but with $\tilde h_{ab}$ on
${\I}$ {\it not} in the equivalence class of the Einstein
static universe.  It would be interesting to extend our analysis
to this case.

Perhaps the most interesting generalization, however, would be to
weaken the asymptotic conditions at infinity and study cases in
which the unphysical metric is less smooth.  While a linearized
analysis indicates that our conditions are consistent with the
dynamics of various fields, it also indicates that other
consistent conditions should exist.  The issue is similar to the
use of Dirichlet, Neumann, or another member of the $S^1$ of
consistent linear boundary conditions for the wave equation on the
half-line. However, an interesting difference here is that some
choices of asymptotic behaviors suggested by the linear analysis
allow the fields to grow at infinity so that the linear
approximation breaks down.

Some progress has recently been made in constructing conserved quantities
under such weakened boundary conditions for scalar fields with masses near
the Breitenlohner-Freedman bound \cite{cecs1,sb1,sb2,ls,sb3,blmsw,sb4}
using the Henneaux-Teitelboim construction.  However, we note that
in $d=4,5,6$ the linearized analysis suggests that even for the case of
pure Einstein-Hilbert gravity one may be able to define a consistent
dynamics (and perhaps conserved charges) with more general asymptotic
conditions. This would be interesting to explore in future work.

\bigskip
\begin{center}
{\bf Acknowledgements}
\end{center}

We thank Thomas Hertog, Gary Horowitz, Harvey Reall, and Bob Wald
for numerous useful discussions.  S.H. and D.M. were supported
in part by NSF grant PHY0354978 and by funds from the University
of California. S.H. was also supported in part by DOE grant DE-FG02-91ER40618.
A.I. was supported in part by JSPS and by NSF grant
PHY 00-90138 to the University of Chicago.

\bigskip

\appendix
\section*{Appendix}

\section{Perturbation equations for the Weyl tensor in pure AdS}

In this section, we give a derivation of eq.~(\ref{eq:eigenvalue})
for the Weyl component $\Psi$. In fact, we will also give a derivation
of corresponding equations in static charts other than the global chart
used in the main part of the paper, namely the chart with
hyperbolic sections, and the chart with flat sections
(the ``horospherical chart''). The latter corresponds to the commonly
used ``Poincare coordinates.'' Some of the material in this section may be
of interest more generally, so we give some detail.

\subsection{Static charts and background quantities}
The metric $g_{ab}$ of pure AdS-space can be written as
\bena
\label{scharts}
 ds^2_0 = - V^2dt^2 + \frac{dr^2}{V^2} + r^2 d\sigma_{(K)}^2 \,,
\eena
where $V=V(r)$ is given as
\ben
  V^2= K + \frac{r^2}{\ell^2} \,,
\een
and $d\sigma_{(K)}^2$ is the metric of a
$(d-2)$-dimensional constant curvature space with the unit sectional
curvature $K=\pm 1,\,0$.
For the global chart $K=1$, for the horospherical chart $K=0$, and
for the hyperbolic chart $K=-1$. Note that the horospherical and
hyperbolic charts only cover a portion of AdS space, and also that
the Killing orbits of $(\partial/\partial t)^a$ are different among
different charts.
The domains covered by the above charts naturally have the structure of
a warped product of an auxiliary 2-dimensional AdS space corresponding
to the $r$-$t$ directions and the $(d-2)$-dimensional constant curvature
space with $K= \pm 1, \, 0$, respectively.

To simplify our notation, we denote by $(y^A, \theta^i) = x^\mu$ the
coordinates adapted to this product structure, i.e.,
\ben
g_{AB}dy^A dy^B = - V^2dt^2 + \frac{dr^2 }{V^2}\,,
\quad
  d\sigma_{(K)}^2 = \sigma_{ij}d\theta^id\theta^j \,.
\een
Thus, the uppercase roman letters
in the range $A, B, \dots $ denote
coordinates of the $2$-dimensonal AdS-space and the lowercase roman
letters in the range $i,j,\dots $ denote the angular
coordinates $\theta^i$ of the $(d-2)$-dimensional space.

Let $\D_A$ and $\hat D_j$ be the derivative operators associated,
respectively, with $g_{AB}$ and $\sigma_{ij}$.
Then the derivative operator $\nabla_a$ associated with $g_{ab}$
is decomposed using the following formulas for the Christoffel symbols:
\bena
 \Gamma^A{}_{ij} = - \frac{\D^A r}{r} g_{ij} = - r\D^A r \sigma_{ij} \,,
\quad
 \Gamma^{i}_{Aj} = \frac{\D_A r}{r}\delta^i{}_j \,,
\label{Christoffel}
\eena
and $\Gamma^A{}_{BC}$ and $\Gamma^i{}_{jk}$ are identical
to the coefficients of $\D_A$ and $\hat D_{j}$, respectively.

\bigskip


Consider a function
\ben
  Y = V(r) \cos (\sqrt{K}t/\ell) \,.
\een
Then the gradient $Z_a = \nabla_a Y$ satisfies
\ben
 \nabla_a Z_b = \frac{Y}{\ell^2} g_{ab} \,.
\een
So, $Z^a $ is a homothety vector.
For later convenience, we list below some formulas:
\bena
&&
 D_A D_B Z_C = \frac{1}{\ell^2}Z_A g_{BC} \,,
\quad
 Z_A Z^A = \frac{1}{\ell^2}(Y^2-K) \,,
\label{formula:A:begin}
\\
&&
 \left(\frac{\dot Y}{Y}\right)^2
 = \frac{K}{\ell^2}\left(\frac{V^2}{Y^2}-1\right) \,, \quad
 \left(V\frac{ Y'}{Y}\right)^2
 = \frac{r}{\ell^2} \,,
\\
&&
\frac{D^Ar}{r} = \frac{\ell^2}{r^2}
                 \left(V^2\frac{Z^A}{Y}+ \frac{\dot Y}{Y}t^A \right) \,,
\quad
D_Ct_A = 2 \frac{Z_{[C}t_{A]}}{Y} \,,
\\
&&
D_A\left(\frac{Z_C}{Y} \right)
= \frac{g_{CA}}{\ell^2}-\frac{Z_C}{Y}\frac{Z_A}{Y} \,,
\quad
\frac{D_C r}{r}\frac{Z^C}{Y}=\frac{1}{\ell^2} \,,
\\
&&
 g_{AB} = 
    \frac{\ell^2}{r^2}\left( \frac{K}{Y^2} - 1 \right)t_At_B
     + V^2\frac{\ell^4}{r^2} \frac{Z_A}{Y} \frac{Z_B}{Y}
     + 2\frac{\ell^4}{r^2}\frac{\dot Y}{Y} \frac{Z_{(A}t_{B)}}{Y} \,,
\label{formula:A:end}
\eena
where the {\em dot} and the {\em prime} denote the partial derivatives
with respect to $t$ and $r$,  respectively, and
$t^a=(\partial /\partial t)^a$ is the background time translation Killing
vector.

\subsection{Weyl curvature perturbations}

Let $\delta C_{abcd}$ denote perturbations of Weyl
curvature on pure AdS-spacetime.
Using the Bianchi identity and the Einstein equations, we have
\bena
 &&
 \left(
      \nabla^e \nabla_e + \frac{2(d-1)}{\ell^2}
 \right)\delta C_{abcd} = 0 \,,
\label{eq:Weyl:start1}
 \\
 &&
\quad
  \nabla^a \delta C_{abcd} = 0 \,.
\label{eq:Weyl:start2}
\eena
We note that $\delta C_{abcd}$ itself is gauge-invariant
since the background Weyl curvatures are all vanishing.

Define
\ben
 \E_{ab} = \delta C_{a c b d} Z^c Z^d \,.
\een
Then,
from eqs.~(\ref{eq:Weyl:start1}) and (\ref{eq:Weyl:start2})
and the homothety property of $Z^a$ one finds that
\bena
 &&
 \left(
      \nabla^c \nabla_c + \frac{2(d-2)}{\ell^2}
 \right)\E_{ab} = 0 \,,
\\
&&
\quad
 \nabla^a \E_{ab} = 0 \,.
\eena
Using the expressions (\ref{Christoffel}) for the Christoffel
symbols $\Gamma^a_{bc}$, we can decompose the above
equations into the following set of evolution equations for the
components $\E_{AB}$, $\E_{Ai}$ and $\E_{ij}$ and constraint equations
among them:
\bena
&&
D^CD_C\E_{AB} + (d-2)\frac{D^Cr}{r} D_C \E_{AB}
-(d-2)\frac{D^Cr}{r}\left(\frac{D_Ar}{r} \E_{BC}
+ \frac{D_Br}{r} \E_{CA}\right)
\nonumber \\
&& \qquad
+ \frac{2(d-2)}{\ell^2} \E_{AB}
+ \frac{\hat \Delta}{r^2}\E_{AB}
\nonumber \\
&& \qquad \quad
- 2\frac{D_Ar}{r}\hat D_m\E^m{}_B
- 2\frac{D_Br}{r}\hat D_m\E^m{}_A
+ 2 \frac{(D_Ar)D_Br}{r^2} \E^m{}_m
= 0 \,,
\\
&&
D^CD_C \E_{Aj} + (d-4)\frac{D^Cr}{r}D_C\E_{Aj}
-\left\{ \frac{D^2r}{r}+(d-3)\frac{(Dr)^2}{r^2}  \right\} \E_{Aj}
\nonumber \\
&& \qquad
- d \frac{(D_Ar)D^Cr}{r^2} \E_{Cj}
+ \frac{\hat \Delta}{r^2} \E_{Aj}
+ \frac{2(d-2)}{\ell^2} \E_{Aj}
\nonumber \\
&& \qquad \quad
- 2\frac{D_Ar}{r}\hat D_m \E^m{}_j
+ 2\frac{D^Cr}{r} \hat D_j \E_{CA}
= 0 \,,
\\
&&
D^CD_C\E_{ij} + (d-6)\frac{D^Cr}{r}D_C\E_{ij}
- 2 \left\{ \frac{D^2r}{r}+(d-4)\frac{(Dr)^2}{r^2}\right\} \E_{ij}
+ \frac{\hat \Delta}{r^2} \E_{ij}
+ \frac{2(d-2)}{\ell^2} \E_{ij}
\nonumber \\
&& \qquad
+ 2\frac{D^Cr}{r}( \hat D_i \E_{Cj} + \hat D_j \E_{Ci})
+ 2 \frac{(D^Cr)D_Ar}{r^2} \E_{CA}g_{ij}
= 0 \,,
\eena
and
\bena
&&
 \D_C\E^C{}_A  + (d-2)\frac{\D^Cr}{r} \E^C{}_A
 + \frac{\D_Ar}{r} \E^C{}_C + \hat D_m \E^m{}_A =0 \,,
\\&&
 \hat D_m\E^m{}_j  + \frac{1}{r^{d-2}} \D_A(r^{d-2}\E^A{}_j)=0 \,,
\eena
where $\hat \Delta = \hat D^m \hat D_m$.

We denote $\E = \E_{AB}t^A t^B$, $\E_i = \E_{Ai}t^A $.
Using formulas listed above (\ref{formula:A:begin})-(\ref{formula:A:end})
and the formulas for $\E,\,\E_i,\,\E_{ij}$ below
\bena
&&
\E_{AB}Z^B = 0 \,,
\\
&&
Z^At^BD_C\E_{AB}  = -\frac{Y}{\ell^2}\E_{BC}t^B \,,
\\
&&
t^At^BD_C\E_{AB} = D_C\E - 2 \frac{Z_C}{Y} \E \,,
\\
&&
Z^AD_C\E_{Aj}= -\frac{Y}{\ell^2}\E_{Cj} \,,
\\
&&
t^AD_C\E_{Aj}=D_C\E_j- \frac{Z_C}{Y} \E_j \,,
\eena
we have evolution equations for $\E,\,\E_i,\,\E_{ij}$
\bena
&&
\left[
  \D^A\D_A +  \left( (d-2)\frac{\D^Cr}{r}- 4\frac{Z^C}{Y}\right) \D_C
  - \frac{6}{\ell^2}\frac{K }{Y^2} + \frac{1}{r^2}\hat \Delta
\right]\E =0 \,,
\label{eq:phiS}
\\
&&
\Bigg[
 \D^A\D_A + \left\{ (d-4)\frac{\D^Cr}{r}- 2\frac{Z^C}{Y}\right\} \D_C
 - \frac{2}{\ell^2}\frac{K }{Y^2}
 - (d-3)\frac{K}{r^2}  + \frac{1}{r^2}\hat \Delta
\Bigg] \E_j
\nonumber \\
&& \quad
 = - 2\frac{\ell^2}{r^2}\frac{\dot Y}{Y} \hat D_j \E \,,
\label{eq:phiV}
\\
&&
 \Bigg[
       \D^A\D_A + (d-6) \frac{\D^Cr}{r}\D_C
       -2(d-4)\frac{K}{r^2} + \frac{1}{r^2}\hat \Delta
 \Bigg] {\cal E}_{ij}
\nonumber \\
&& \quad
 =
- 2\frac{\ell^2}{r^2} \frac{\dot Y}{Y}(\hat D_i \E_j+\hat D_j\E_i)
- 2\left(\frac{\ell^2}{r^2}\frac{\dot Y}{Y} \right)^2 g_{ij} \E \,.
\label{eq:phiT}
\eena
{}From the constraint equations, we obtain
\bena
&&
 \frac{1}{V^2}\frac{r^2}{\ell^2}
  \left(
          -t^C + \ell^2\frac{\dot Y}{Y}\frac{D^Cr}{r}
  \right) D_C \E + (d-3)\frac{\dot Y}{Y}\E
 + \frac{1}{\ell^2} \hat D^m\E_m
 = 0 \,,
\label{constraint:sv}
\\
&&
 \frac{1}{V^2}\frac{r^2}{\ell^2}
   \left(
         -t^C + \ell^2\frac{\dot Y}{Y}\frac{D^Cr}{r}
   \right)D_C \E_j
 + (d-3)\frac{\dot Y}{Y}\E_j
 + \frac{1}{\ell^2}\hat D^m {\cal E}_{mj}
 = 0 \,.
\label{constraint:vt}
\eena

\subsection{Master equation}
Although the equations for $\E$, $\E_i$, $\E_{ij}$ obtained above
are coupled each other, utilizing the symmetry of the $(d-2)$-dimensional
constant curvature space, we can obtain a set of decoupled equations.
To do this, we decompose $\E_i$ into a scalar component and a divergence-free
vector component with respect to $d\sigma_K^2$, and similary,
decompose $\E_{ij}$ into a scalar, a divergence-free vector, and
a transverse-traceless tensor component with respect to $d\sigma_K^2$.
It is convenient to introduce harmonic functions ${\Bbb S}_{\bf k}$,
vectors ${\Bbb V}_{\bf k}{}_i$, and symmetric tensors
${\Bbb T}_{\bf k}{}_{ij}$ on $\sigma_{ij}{}_{K}$ defined by
the eigenvalue equations,
\bena
&&
 (\hat \Delta + {\bf k}_S^2) {\Bbb S}_{\bf k}{} = 0 \,,
\\
&&
 (\hat \Delta + {\bf k}_V^2) {\Bbb V}_{\bf k}{}_i = 0 \,,
\quad \hat D_i {\Bbb V}_{\bf k}{}^i = 0 \,,
\\
&&
 (\hat \Delta + {\bf k}_T^2) {\Bbb T}_{\bf k}{}_{ij} = 0 \,,
 \quad {\Bbb T}_{\bf k}{}^i{}_i = 0 \,,
 \quad \hat D_i {\Bbb T}_{\bf k}{}^i{}_j = 0 \,.
\eena
Some properties of these tensor harmonics and decomposition theorems
are given in \cite{iw,KIS2000}.
We then expand $\E$, $\E_i$, $\E_{ij}$ in terms of these harmonics
as follows:
\bena
\E &=& \psi_S \scalar \,,   
\\
\E_i &=& \phi_S \hat D_i \scalar 
+ \psi_V \vector_i \,, 
\\
\E_{ij} &=& \E_L \sigma_{ij}\scalar
+ \E_T \left(\hat D_i \hat D_j - \frac{1}{d-2}\hat \Delta \sigma_{ij}\right)
\scalar
+ \E_V \hat D_{(i}\vector_{j)}
+ \psi_T \tensor_{ij} \,, 
\eena
where here and hereafter we omit the suffix ${\bf k}$ labeling
the eigenvalue of each mode and the mode summation
symbol $\sum_{\bf k}$ over them.
We call $(\psi_S,\,\phi_S,\,\E_L,\,\E_T)$ the scalar-type components,
$(\psi_V,\,\E_V)$ the vector-type components, and $\psi_T$ a tensor-type
component of the Weyl perturbations. Note that the tensor-type component
does not exist for $d=4$ case.

The expansion coefficients are not independent due to the constraint
equations~(\ref{constraint:sv}) and (\ref{constraint:vt}).
First, the tracelessness $\E^a{}_a =0$ implies that $\E_L$ is
described by $\psi_S$.
It follows from eq.~(\ref{constraint:sv}) that $\phi_S$ is described
in terms of $(\psi_S, \,\partial_t \psi_S, \, \partial_r \psi_S)$,
and from eq.~(\ref{constraint:vt}) that $\E_T$ is described in terms of
$(\phi_S, \,\partial_t \phi_S, \, \partial_r \phi_S)$ and $\E_V$
in terms of $(\psi_V, \,\partial_t \psi_V, \, \partial_r \psi_V)$.
Therefore once we obtain $\psi_S$, $\psi_V$, and $\psi_T$, we in fact obtain
all of the components.

Furthermore, since the tensor harmonics $\tensor_{ij}$ have $d(d-4)/2$
independent components, the vector harmonics $\vector_i$ have $d-3$
independent components, and the scalar harmonics $\scalar$ have one
independent component, one finds the total number of independent
components $d(d-3)/2$, which corresponds to the number of dynamical
degrees of freedom for gravitational radiation in $d$-dimensional spacetime.
Therefore $\psi_S$, $\psi_V$, and $\psi_T$ describe all the dynamical
modes of gravitational perturbations.

Now we derive below a master equation that governs
$\psi_S$, $\psi_V$, $\psi_T$, and hence all dynamical degrees of freedom
of gravitational perturbations. We can express
equations~(\ref{eq:phiS}), (\ref{eq:phiV}), (\ref{eq:phiT})
as decoupled equations, respectively, for $\psi_S$, $\psi_V$, $\psi_T$:
\bena
&&
 \left(
       \frac{\partial^2}{\partial t^2}
      - 4\frac{\dot Y}{Y}\frac{\partial}{\partial t}
      + 6\frac{K}{\ell^2}\frac{V^2}{Y^2}
 \right) \psi_S
 \non \\
 && \qquad \quad
 = \left\{
            \frac{V^2}{r^{d-2}} \frac{\partial}{\partial r}
            \left( r^{d-2} V^2\frac{\partial}{\partial r} \right)
            - 4 V^2 \frac{Y'}{Y} V^2\frac{\partial}{\partial r}
            - \frac{V^2}{r^2}{\bf k}_S^2
     \right\} \psi_S \,,
\\
&&
 \left(
       \frac{\partial^2}{\partial t^2}
      - 2\frac{\dot Y}{Y}\frac{\partial}{\partial t}
      + 2\frac{K}{\ell^2}\frac{V^2}{Y^2}
 \right) \psi_V
  = \left[
          \frac{V^2}{r^{d-4}} \frac{\partial}{\partial r}
          \left( r^{d-4} V^2\frac{\partial}{\partial r} \right)
    \right.
\non \\
&& \,\qquad \qquad
    \left.
          - 2 V^2\frac{Y'}{Y} V^2\frac{\partial}{\partial r}
          - \left\{(d-3)K + {\bf k}_V^2 \right\} \frac{V^2}{r^2}
    \right] \psi_V \,,
\\
&&
\frac{\partial^2}{\partial t^2} \psi_T
= \left[
         \frac{V^2}{r^{d-6}} \frac{\partial}{\partial r}
         \left( r^{d-6} V^2\frac{\partial}{\partial r} \right)
         - \left\{ 2(d-4) K +  {\bf k}_T^2 \right\} \frac{V^2}{r^2}
    \right] \psi_T \,.
\eena
Define master variables $\Psi_S$, $\Psi_V$, and $\Psi_T$ by
\ben
  \psi_S = Y^2r^{-(d-2)/2}\Psi_S \,, \quad
  \psi_V = Y r^{-(d-4)/2}\Psi_V \,, \quad
  \psi_T = r^{-(d-6)/2}\Psi_T \,,
\label{def:new-master-variables}
\een
and introduce the following radial function $x$;
\bena
  x = -\frac{1}{\ell} \int \frac{dr}{V} \,,
\quad \mbox{so that}\quad
\frac{r}{\ell} = \frac{\sqrt{K}}{\tan(\sqrt{K} x)} \,,
\eena
thus $x=0$ at infinity.
Then, one finds that the three equations for $\Psi_S$, $\Psi_V$, $\Psi_T$
are expressed exactly in the same form
\bena
 \ell^2\frac{\partial^2}{\partial t^2} \Psi
 = \left\{
        \frac{\partial^2}{\partial x^2}
     -  \left(\nu^2-\frac{1}{4} \right) \frac{K}{\sin^2 (\sqrt{K}x)}
     -  \left(\sigma^2-\frac{1}{4}\right) \frac{1}{\cos^2 (\sqrt{K}x)}
   \right\} \Psi \,,
\label{eq:master:weyl}
\eena
where we have denoted $\Psi_S$, $\Psi_V$, and $\Psi_T$ universally by $\Psi$.
Here it is understood that
$\sin(\sqrt{K}x)/\sqrt{K}=x$ and $\cos(\sqrt{K}x)=1$ for $K=0$.
The second term of the right-hand side of eq.~(\ref{eq:master:weyl})
becomes relevant near infinity
(where $x=0$) and the parameter $\nu$ depends only on the spacetime
dimension:
\ben
  \nu^2-\frac{1}{4} = \frac{(d-4)(d-6)}{4} \,.
\een
Without loss of generality, we take $\nu$ to be non-negative,
\ben
\nu = \frac{|d-5|}{2} \,.
\een
The third term on the right-hand side of eq.~(\ref{eq:master:weyl})
stems from to the angular-momentum. The parameter $\sigma$
depends on the spacetime dimension, the sectional curvature $K$ of
$d \sigma^2$ and the mode of perturbations,
\ben
  \sigma^2-\frac{1}{4} = \frac{(d-2)(d-4)}{4}K + {\bf k}_K^2 \,,
\label{def:sigma}
\een
where ${\bf k}_K^2= {\bf k}_S^2$ for the scalar-type,
${\bf k}_K^2= {\bf k}_V^2+K$ for the vector-type, and
${\bf k}_K^2= {\bf k}_T^2+2K$ for the tensor-type perturbation.
In particular, for $K=1$, the eigenvalues are ${\bf k}_S^2= l(l+d-3)$,
${\bf k}_V^2= {\bf k}_S^2-1$,
${\bf k}_T^2= {\bf k}_S^2-2$, hence in this case
${\bf k}_K^2 = l(l+d-3)$ for all types of perturbations.
Therefore, in this case, one can universaly take
\ben
\sigma = l + \frac{d-3}{2} \,.
\een

Since near infinity $Y \sim r/\ell$,
the definition eq.~(\ref{def:new-master-variables}) and
the master equation~(\ref{eq:master:weyl}) imply that
in the asymptotic region, all the tensorial types of perturbations
behave in the same way
\ben
\psi_{S,V,T} \sim r^{-(d-6)/2}\Psi \,.
\een


%


\medskip
For concreteness, let us examine the master equation~(\ref{eq:master:weyl})
in the various charts.

\medskip
\noindent
{\bf The global chart:} In this chart, $K=1$ and
the radial coordinates $r$ and $x$ are related by
\ben
 \frac{r}{\ell} = \frac{\cos x}{\sin x} \,.
\een
The master wave equation (\ref{eq:master:weyl}) becomes
\ben
  \left(
       - \frac{\partial^2}{\partial x^2}
       + \frac{\nu^2 - 1/4}{\sin^2 x}
       + \frac{\sigma^2-1/4}{\cos^2 x}
  \right) \Psi =  \omega^2 \Psi \,,
\label{eq:eigenvalueK=1}
\een
where the eigenvalue $\omega^2$ satisfies
$ \ell^2 \partial^2 \Psi/\partial t^2 = -\omega^2 \Psi$.
Note that $\omega$ can be an arbitrary complex number if no
conditions on the type of solutions are imposed.
Note also that in the $K=1$ case, ${\bf k}^2 =l(l+d-3)$ takes on
discrete values, and the same is consequently true for
$\sigma = l + (d-3)/2$, with $l=2,\,3,\, \dots$.
The solutions to this equation are given by hypergeometric functions
but the only solution that is regular at the center of
$(S^{d-2}, d\sigma_{+1}^2)$ is given by eq.~(\ref{sol:smooth}).

\medskip
\noindent
{\bf Horospherical chart:}
In this chart, $K=0$ and we have
\ben
\frac{r}{\ell} = \frac{1}{x},
\een
and the wave equation~(\ref{eq:master:weyl}) becomes
\ben
  \left(
       - \frac{\partial^2}{\partial x^2}
       + \frac{\nu^2 - 1/4}{x^2}
       + {\bf k}^2
  \right) \Psi =  \omega^2 \Psi \,,
\label{eq:eigenvalue1}
\een
where $\omega, {\bf k}$ are real numbers, i.e., in the $K=0$ case
${\bf k}^2$ is in the continuous spectrum. The solutions to this equation
are given in terms of Bessel and Neumann functions, i.e.,
$\Psi \propto C_1 \sqrt{x}J_\nu(\sqrt{\omega^2-{\bf k}^2}x)
+ C_2 \sqrt{x}N_\nu(\sqrt{\omega^2-{\bf k}^2}x)$.

\medskip
\noindent
{\bf Hyperbolic chart:}
In this chart, $K=-1$ and we have
\ben
  \frac{r}{\ell} = \frac{\cosh x}{\sinh x} \,.
\een
In this chart, there exists an event horizon with respect to the Killing
orbits of $(\partial /\partial t)^a$ at $x \rightarrow \infty$.
In particular, if one considers a quotient of the background AdS-space
by a certain discrete subgroup of the hyperbolic isometries of
$d\sigma_{-1}^2$, the resultant spacetime describes a BTZ-type black
hole spacetime.
The wave equation~(\ref{eq:master:weyl}) becomes
\ben
  \left(
       - \frac{\partial^2}{\partial x^2}
       + \frac{\nu^2 - 1/4}{\sinh^2 x}
       + \frac{\sigma^2-1/4}{\cosh^2 x}
  \right) \Psi =  \omega^2 \Psi \,.
\een
The solutions to this equation may be given in terms of
hypergeometric functions in an analogous manner to the
case (\ref{eq:eigenvalueK=1}).
In this case, however, $\sigma^2 -1/4$ given by eq.~(\ref{def:sigma})
can take a negative value for the higher dimensional case $d \geq 5$,
and also $x \rightarrow \infty$ corresponds to the event horizon.
Therefore one might need to consider with  more care the regularity of solutions
at $x \rightarrow \infty$.

\section{Perturbation equations for anti-symmetric tensor fields in pure AdS}

This appendix analyzes the behavior of anti-symmetric tensor
fields $A_{a_1 \ldots a_p}$ satisfying equations of motion of the form
\ben
\label{AEOM}
\nabla^b F_{b a_1 \ldots a_p} = -m^2 A_{a_1 \ldots a_p},
\een
where $F = dA$ and $p+1 \le n$. Here we take the fields to propagate
in pure anti-de Sitter space. The kinetic operator on the left-hand side
is referred to as the Maxwell operator in \cite{AdS5,AdS5krn,AdS7}.
These references show that the linearized anti-symmetric tensor fields
that arise in reductions of 10- and 11-dimensional supergravity to $AdS_5$
and $AdS_7$ satisfy equations of motion of this type with $m^2 \ge 0$.

It is convenient to take the exterior derivative of (\ref{AEOM}).  In doing so, it is useful to note the following relation:
\ben
\nabla_{[c_1} \nabla^a \nabla_b A_{c_2 \ldots c_{p+1}]} =
\nabla_{[c_1} \nabla^a F_{b c_2 \ldots c_{p+1}]} =
- \nabla_{[b} \nabla^a F_{c_1 c_2 \ldots c_{p+1}]}.
\een
Thus one finds
\bena
-(p+2) m^2 F_{c_1 c_2 \ldots c_{p+1}} &=& - \nabla_a \nabla^a F_{c_1 c_2 \ldots c_{p+1}} + (p+1) \nabla_{[c_1| } \nabla^a F_{a| c_2 \ldots c_{p+1}]} \cr &=&
- \nabla_a \nabla^a F_{c_1 c_2 \ldots c_{p+1}} -m^2  (p+1)  F_{c_1 c_2 \ldots c_{p+1}}.
\eena
That is,
\ben
\nabla_a \nabla^a F_{c_1 c_2 \ldots c_{p+1}} =  m^2 F_{c_1 c_2 \ldots c_{p+1}}.
\een

It is straightforward to work out these equations in detail in the $K=0$ static chart of equation (\ref{scharts}).  Introducing $z=  1/r$,  we find
\bena
\label{eEOM}
 m^2 F_{\hat c_1 \ldots \hat c_{p+1}} &=&  -z^2 (- \partial_t^2 + \sigma^{ij} \hat D_i \hat D_j ) F_{\hat c_1 \ldots \hat c_{p+1}} \cr &-& z(2p + 2 - d) \partial_z F_{\hat c_1 \ldots \hat c_{p+1}}  + (p+1)(p+1-d) F_{\hat c_1 \ldots \hat c_{p+1}}, \ \ \ \
\eena
where the hats on the indices $\hat c_i$ indicate that we have chosen
$\hat c_i \neq z$.
Defining $F_{\hat c_1 \ldots \hat c_{p+1}} = z^\alpha \chi$
with $\alpha = (-1+d-2p)/2$ and Fourier transforming on ${\I}$
then places (\ref{eEOM}) in the standard form of Bessel's equation.
Thus one finds that the two indepdendent solutions are Bessel functions
whose leading behavior at small $z$ is $z^\nu$ with
\ben
\label{nu}
\nu = \pm \sqrt{m^2 + \frac{d(d+1)}{4} - \frac{1+p}{2} }.
\een
Since $d \ge p+1 $ and $d \ge 3$, the argument of this square root is bounded below by $m^2+ 3/2$.
{}From (\ref{AEOM}), we then see that the $z$-components of $F$ must fall-off at least as fast.

We now require (in analogy with \cite{iw})  that boundary conditions be imposed such that the time-translation operator is self-adjoint with respect to the natural inner product
\ben
\label{asip}
\int_\Sigma \sqrt{g_{\Sigma}} V^{-1} g^{a_1b_1} \ldots g^{a_{p+1} b_{p+1}} F_{a_1\ldots a_{p+1}}
F_{b_1\ldots b_{p+1}},
\een
where $\Sigma$ is a Cauchy surface and $V^2 = z^{-2}$ is the norm of the time translation $\partial_t$.
This will be the case if eigenfunctions of $\partial_t$ satisfying the boundary condition are normalizable in the inner product (\ref{asip}).  Since we are interested in the case $m^2 \ge 0$, we have $|\nu| \ge 1$ which requires us to choose the positive branch of (\ref{nu}).  In particular, we find  $\nu \ge \sqrt{3/2}$.  As a result, one may show that the anti-symmetric tensor fields fall off too fast at infinity to contribute to any expression for the conserved energy.


\end{document}